\documentclass[aps,prd,twocolumn,superscriptaddress,amsfont,amssymb,amsmath,nofootinbib,showpacs,balancelastpage]{revtex4-1}
\usepackage{float}
\usepackage{epsfig}
\usepackage{graphicx}
\usepackage{dcolumn}
\usepackage{bm}
\usepackage{color}
\usepackage{tabularx}
\usepackage{multirow} 
\usepackage{booktabs} 
\usepackage{tabularx} 
\usepackage{cleveref}
\usepackage{hyperref,xcolor}
\hypersetup{
  colorlinks,
  citecolor=blue,
  linkcolor=blue,
  urlcolor=black}
\usepackage{xcolor}
\usepackage{epstopdf}

\newcommand{\Od}{{\mathcal O}}

\def\thebiblio#1{
\begin{center}\bf \large References
\end{center}
\list
{[\arabic{enumi}]}{\settowidth\labelwidth{#1.}\leftmargin\labelwidth
 \advance\leftmargin\labelsep
 \usecounter{enumi}}
 \def\newblock{\hskip .11em plus .33em minus -.07em}
 \sloppy
 \sfcode`\.=1000\relax}

\begin{document}

\preprint{}
\title{Finite temperature corrections to the energy-momentum tensor\\ at one-loop in static space-times}

\author{Franco D.\ Albareti}
\email{`la Caixa'-Severo Ochoa Scholar, franco.albareti@csic.es}
\affiliation{Instituto de F{\'i}sica Te{\'o}rica UAM/CSIC, Universidad Aut{\'o}noma de Madrid, Cantoblanco, E-28049 Madrid, Spain}
\affiliation{Campus of International Excellence UAM+CSIC, Cantoblanco, E-28049 Madrid, Spain}

\author{Antonio L.\ Maroto}
\email{maroto@ucm.es}
\affiliation{Departamento de F\'{\i}sica Te\'orica and UPARCOS, Universidad Complutense de Madrid, 28040 
Madrid, Spain}

\author{Francisco Prada}
\email{f.prada@csic.es}
\affiliation{Instituto de F{\'i}sica Te{\'o}rica UAM/CSIC, Universidad Aut{\'o}noma de Madrid, Cantoblanco, E-28049 Madrid, Spain}
\affiliation{Campus of International Excellence UAM+CSIC, Cantoblanco, E-28049 Madrid, Spain}
\affiliation{Instituto de Astrof{\'i}sica de Andaluc{\'i}a (CSIC), Glorieta de la Astronom{\'i}a, E-18080 Granada, Spain}

\date{\today}


\begin{abstract}

Finite temperature corrections to the effective potential and the  energy-momentum tensor of a scalar field are computed in a perturbed Minkoswki space-time. We consider the explicit mode decomposition of the field in the perturbed geometry and obtain   analytical expressions in the non-relativistic and ultra-relativistic limits to first order in scalar metric perturbations. In the static case, our results are in agreement with previous calculations based on the Schwinger-De Witt expansion which indicate that  thermal effects in a curved space-time can be encoded in the local Tolman temperature at leading order in perturbations and in the adiabatic expansion. We also study the shift of the effective potential minima produced by thermal corrections in the presence of static gravitational fields. Finally we discuss the dependence on the initial conditions set for the mode solutions.

\end{abstract}


\maketitle

\onecolumngrid


\section{Introduction}


Finite-temperature corrections to the effective potential in quantum field theory 
play a fundamental role in the description of phase transitions in the 
early universe. In particular, symmetry restoration at high temperatures is
an essential ingredient of the Higgs mechanism for electroweak symmetry breaking.
 In flat space-time, such thermal corrections were computed for the 
first time 
in the seminal papers of Dolan and Jackiw \cite{DJ} and Weinberg \cite{Weinberg} using thermal Green functions methods. The possibility of extending those methods to 
more realistic scenarios incorporating space-time curvature 
meets certain difficulties since finite temperature field 
theory is only well-defined provided the geometry possesses a global time-like Killing field.
Thus for example, for static or stationary space-times, the thermal Green functions method has been
applied for homogeneous and isotropic  Einstein static spaces in \cite{DC}. These methods were extended to conformally static Robertson-Walker backgrounds in \cite{Dr}. The conditions for the construction of a thermal field theory in more general expanding universes (not strictly static) were discussed in \cite{Parker,Hu} where the adiabatic techniques were introduced.

An alternative approach to the adiabatic expansion for thermal field theory in general curved
space-time is the so called Schwinger-DeWitt \cite{Schwinger,DeWitt} expansion of the effective action.
Both approaches are known to agree in the results for the ultraviolet divergences 
in zero-temperature field theory. The Schwinger-DeWitt expansion, being a local curvature expansion, is manifestly covariant but it is not sensitive to the global properties of the space-time such as the presence of boundaries and does not contain information about the 
non-local part of the effective action. Going beyond the Schwinger-DeWitt approximation
requires brute force methods based on explicit mode summation \cite{Birrell}.
Thus for example, in \cite{Huang} phase transitions in homogeneous but anisotropic Bianchi I  and Kasner cosmologies 
were studied using explicit modes sum. 
 In recent works \cite{Maroto, Higgs}, we started this program in the 
 case of weak inhomogeneous gravitational fields by studying the one-loop corrections to the vaccum expectation value (VEV) of the energy-momentum tensor and the effective potential of a massive scalar field. Thus, in \cite{Maroto}, using a regularization procedure based on a simple comoving cutoff, a nonvanishing contribution of metric perturbations to the effective potential was obtained. However, the renormalization
 procedure required the use of non-covariant counterterms. In contrast, dimensional regularization was used in \cite{Higgs} to isolate the divergences, applying techniques developed specifically to deal with non-rational integrands. In this case,  the renormalized
 effective potential, being explicitly covariant, did not contain contributions from the 
 inhomogeneous  gravitational fields  at the leading order in metric perturbation and in the adiabatic expansion in both static and cosmological space-times. 

In this work, we extend these methods to include  finite temperature effects. The inclusion of the Bose-Einstein factor, accounting for the statstical distribution of the energy states, produces a smooth behavior of all quantities involved at large energies, not being necessary to apply any regularization technique (once the vacuum contribution is renormalized). As mentioned above,  in order to compute the aforementioned contribution, we apply the "brute force" method described in \cite{Maroto, Higgs}, i.e.\ performing a summation over the perturbed modes of the quantum field obtained as solutions of the Klein-Gordon equation. We are able to get analytical expressions for the effective potential and the energy-momentum tensor in the non-relativistic and the ultra-relativistic limits. In the static limit, we find that local gravitational effects can be taken into account through the Tolman temperature \cite{Tolman}. This is in accordance with computations of the energy-momentum tensor of a scalar field at finite temperature in a static space-time using the Schwinger-DeWitt approach, \cite{Nakazawa} and \cite{Holstein}. However, we also obtain the explicit time dependence of the expectation values for finite times, which shows that the Tolman temperature can only be defined in the asymptotic time regions.

The work is organized as follows. Section \ref{sec:back} describes the general approach to compute an expectation value over a thermal state in a perturbed FRW metric. The particular expressions to be computed in the case of static space-times are presented in Section \ref{sec:static}. Section \ref{sec:low} and \ref{sec:high} explain the approximations applied to obtain the final result in the non-relativitic and ultra-relativistic limits, respectively. Shifts in the minimum of the effective potential produced by thermal correction are discussed in Section \ref{sec:shift}. Our conclusions are presented in Section \ref{sec:con}.


\section{Finite temperature corrections}
\label{sec:back}

Given a scalar field $\phi$, with potential $V(\phi)$, its classical action in a $(D+1)$-dimensional space-time with 
metric tensor $g_{\mu\nu}$ can be written as
\begin{eqnarray}
S[\phi,g_{\mu\nu}]\,=\,\int \text{d}^{D+1}x \,\sqrt{g} \left(\frac{1}{2}\,g^{\mu\nu}\,\partial_\mu\phi\,\partial_\nu\phi\,-\,V(\phi)\right).
\end{eqnarray}
%
%
%
As is well known, the solutions $\phi=\hat \phi$ of the classical equation of motion
\begin{eqnarray}
\Box \, \hat\phi \,+\,V'(\hat\phi)\, =\, 0 \,
\label{KGc}
\end{eqnarray}
are those that minimize the action. On the other hand, quantum fluctuations around the classical solution
$\delta \phi=\phi-\hat\phi$ satisfy the equation of motion
\begin{eqnarray}
\left(\Box \, +\,m^2(\hat\phi)\right)\delta\phi\, =\, 0
\label{pKGc}
\end{eqnarray}
with
\begin{eqnarray}
m^2(\hat\phi)\,=\,V''(\hat\phi)\,.
\end{eqnarray}

Let us consider a metric which can be written as a scalar perturbation around  
 a flat Robertson-Walker background
\begin{eqnarray}
\text{d}s^2 \,=\,a^2(\eta) \left\{ \left[1 + 2 \Phi(\eta,{\bf x})\right]\, \text{d}\eta^2 - \left[1 - 2\Psi(\eta,{\bf x})\right]\,\text{d}{\bf x}^2 \right\}\label{metric}
\end{eqnarray} 
where $\eta$ is the conformal time, $a(\eta)$ the scale factor, and $\Phi$ and $\Psi$ are the scalar perturbations in the longitudinal gauge. Given this geometry, the mode  solutions $\delta\phi_k$ to \eqref{pKGc} can be found using a WKB approximation to first order in metric perturbations and to the leading adiabatic order as \cite{Higgs}
\begin{eqnarray}
\delta\phi_k(\eta,{\bf x})\,=\,\delta\phi_k^{(0)}(\eta,{\bf x})\left(1\,+\,P_k(\eta,{\bf x})\,+\,i\,\delta\theta_k(\eta,{\bf x})\vphantom{\frac{1}{1}}\right)
\label{sol}
\end{eqnarray}
where
\begin{eqnarray}
\delta\phi_k^{(0)}(\eta,{\bf x})\,=\,\frac{1}{(2\,\pi)^{D/2}}\frac{1}{a(\eta)^{(D-1)/2}\,\sqrt{2\,\omega_k}}\,e^{i{\bf k}\cdot{\bf x}-i\int^\eta\omega_k(\eta')\text{d}\eta'}\,;
\label{sol0}
\end{eqnarray}
are the unperturbed mode solutions with
\begin{eqnarray}
\omega_k^2(\eta)=k^2+m^2a^2(\eta)
\end{eqnarray}

 The explicit expressions for $P_k(\eta,{\bf x})$ and $\delta\theta_k(\eta,{\bf x})$ in Fourier space are shown in Appendix \ref{appsol}.

The effects of quantum fluctuations on the classical field configuration can be taken into account using the one-loop effective potential \cite{Mukhanov,Higgs}
\begin{eqnarray}
V_{\text{eff}}(\hat\phi)\,=\,V(\hat\phi)\,+\,\frac{1}{2}\int_0^{m^2(\hat\phi)} \text{d}m^2 \langle \delta\phi^2 \rangle\,
\label{effpot}
\end{eqnarray}
where $V(\hat\phi)$ is the tree-level potential and the expectation value of the operator $\langle \delta\phi^2 \rangle$ is taken over a particular quantum state of the field. Taking into account \eqref{sol} and \eqref{sol0} and assuming that the quantum state has a fixed number of particles per mode $n_k$, the one-loop contribution to the effective potential reads
\begin{eqnarray}
\label{effpotT}
\frac{1}{2}\int_0^{m^2(\hat\phi)} \text{d}m^2 \;\langle \delta\phi^2 \rangle\,&=&\,\frac{1}{(2\pi)^D\,a^{D-1}(\eta)}\frac{1}{2}\int_0^{m^2(\hat\phi)} \text{d}m^2 \,\int\text{d}^D{\bf k}\left(\frac{1}{2}\,+\,n_k\right)\frac{1\,+\,2\,P_k(\eta,{\bf p})}{\sqrt{k^2\,+\,m^2\,a^2(\eta)}}\\
&=&\,\frac{2}{(2\pi)^D\,a^{D-1}(\eta)}\,\frac{2\,\pi^{(D-1)/2}}{\Gamma((D-1)/2)}\frac{1}{2} \int_0^{m^2(\hat\phi)} \text{d}m^2 \,\int_0^{\infty}\text{d}k\,k^{D-1}\left(\frac{1}{2}\,+\,n_k\right)\frac{1\,+\,\hat{P}_k(\eta,{\bf p})}{\sqrt{k^2\,+\,m^2\,a^2(\eta)}}\,\nonumber
\end{eqnarray}
where we have defined
\begin{eqnarray}
\hat{P}_k (\eta,{\bf p})\,=\,\int_{-1}^1 \text{d}\hat{x}\,\left(1-\hat{x}^2\right)^{(D-3)/2}\,P_k (\eta,{\bf p})\,,
\end{eqnarray}
being $\hat{x}={\bf k}\cdot {\bf p}/(k\,p)$ and including the general integration measure in $D$ dimensions.

From now on, we consider a thermal quantum state. Then, the number of particles per mode is given by the Bose-Einstein distribution
\begin{eqnarray}
n^T_{k}\,=\,\frac{1}{e^{\omega_k/T}-1}\,,
\end{eqnarray}
where $T$ is the temperature of the state, for the moment understood as a parameter of the Bose-Einstein distribution (see next section).

Let us define $V_1(\hat\phi)$ as the one-loop quantum vacuum contribution, i.e.
\begin{eqnarray}
V_1(\hat\phi)&=&\,\frac{1}{(2\pi)^D\,a^{D-1}(\eta)}\,\frac{2\,\pi^{(D-1)/2}}{\Gamma((D-1)/2)}\, \int_0^{m^2(\hat\phi)} \text{d}m^2 \,\int_0^{\infty}\text{d}k\,k^{D-1}\frac{1}{2}\frac{1\,+\,\hat{P}_k(\eta,{\bf p})}{\sqrt{k^2\,+\,m^2\,a^2(\eta)}}\,
\end{eqnarray}
and $V_T(\hat\phi)$ as the term that includes finite temperature corrections
\begin{eqnarray}
V_T(\hat\phi)&=&\,\frac{1}{(2\pi)^D\,a^{D-1}(\eta)}\,\frac{2\,\pi^{(D-1)/2}}{\Gamma((D-1)/2)} \int_0^{m^2(\hat\phi)} \text{d}m^2 \,\int_0^{\infty}\text{d}k\,k^{D-1}\,n_k^T\,\frac{1\,+\,\hat{P}_k(\eta,{\bf p})}{\sqrt{k^2\,+\,m^2\,a^2(\eta)}}\,
\end{eqnarray}
so that we can write the one-loop effective potential at 
finite temperature as
\begin{eqnarray}
V_{\text{eff}}(\hat\phi)\,=\,V(\hat\phi)\,+\,V_1(\hat\phi)\,+\,V_T(\hat\phi)\,.
\end{eqnarray}

It is important to notice that both the vacuum and the thermal contributions have a homogeneous term, corresponding to the background geometry, and an inhomogeneous one, proportional to the perturbations. Then,
\begin{eqnarray}
V_1({\hat \phi})\,=\,V^{\text{h}}_1({\hat \phi})\,+\,V^{\text{i}}_1({\hat \phi})\,,\\
V_T({\hat \phi})\,=\,V^{\text{h}}_T({\hat \phi})\,+\,V^{\text{i}}_T({\hat \phi})\,.
\end{eqnarray}
The homogeneous part due to vacuum effects $V_1^{\text{h}}({\hat \phi})$, after applying the minimal substraction scheme $\overline{\text{MS}}$ in dimensional regularization with $D=3+\epsilon$, is given by \cite{Higgs}
\begin{eqnarray}
V^{\text{h}}_1(\hat\phi)\,=\,
\frac{m^4(\hat\phi)}{64\pi^2}\left[\ln\left(\frac{m^2(\hat\phi)}{\mu^2}\right)-\frac{3}{2}\right].
\end{eqnarray}
A detailed analysis of the inhomogeneous part of the vacuum $V_1^{\text{i}}({\hat \phi})$ was performed in \cite{Maroto} with a cutoff regularization, and in \cite{Higgs} using dimensional regularization. When a cutoff $\Lambda$ is used, the 
result turns out to be proportional to $m^2(\hat\phi)\Lambda^2\Phi$ in the static case, 
i.e. only the quadratic divergence appears.  
In dimensional regularization we find to first order in perturbations and to the leading adiabatic order that
\begin{eqnarray}
V^{\text{i}}_1(\hat\phi)\,=\,0
\end{eqnarray}
in agreement with the absence of logarithmic divergences in the cutoff case.

In this work, we focus on the thermal contribution $V_T(\hat\phi)$. The corresponding  
inhomogeneous contribution can  in turn be split in the terms proportional to  
$\Phi$ and $\Psi$ as
\begin{eqnarray}
V_T(\hat\phi)\,&=&\,V^{\text{h}}_T(\hat\phi)\,+\,V^\Phi_T(\hat\phi)\,+\,V^\Psi_T(\hat\phi)
\end{eqnarray}
It is important to note that  expression \eqref{effpot} defines the potential except for the addition of an arbitrary function which could depend on the space-time coordinates and the temperature. This function does not modify the dynamics of the field \eqref{KGc}  since it does not introduce any dependence on $m(\hat\phi)$.

In the same fashion, the thermal contribution to the components of the energy-momentum tensor can be obtained from the expressions given in the reference \cite{Higgs} including the number of particles per mode $n_k^T$, thus
\begin{eqnarray}
\langle T^0_{\;0} (\eta,{\bf p})\rangle\,&=&\,\rho(\eta,{\bf p})\,=\,\frac{1}{(2\pi)^D}\frac{1}{a^{D+1}}\int \text{d}^D {\bf k}\,\left(\frac{1}{2}\,+\,n_k^T\right)\omega_k\left[1\,+\,2\,\frac{k^2}{\omega_k^2}\,\Psi({\bf p})\,+\,2\,P_{k}(\eta,{\bf p})\,+\,2\,i\,\frac{{\bf k\cdot p}}{\omega_k^2}\,\delta\theta_k(\eta,{\bf p})\right]\\
\langle T^i_{\;i}(\eta,{\bf p})\rangle\,&=&\,-p_i(\eta,{\bf p})\,=\,-\frac{1}{(2\pi)^D}\frac{1}{a^{D+1}}\int \text{d}^D\, {\bf k}\,\left(\frac{1}{2}\,+\,n_k^T\right)\left[\frac{k^2_i}{\omega_k}\left(1\,+\,2\,\Psi({\bf p})\,+\,2\,P_{k}(\eta,{\bf p})\vphantom{\frac{1}{1}}\right)\,+\,2\,i\,\frac{k_i\,p_i}{\omega_k}\,\delta\theta_k(\eta,{\bf p})\right]\nonumber\\ \ \\
\langle T^i_{\;0}(\eta,{\bf p})\rangle\,&=&\,\frac{1}{(2\pi)^D}\frac{1}{a^{D+1}}\int \text{d}^D {\bf k}
\left(\frac{1}{2}\,+\,n_k^T\right)\left[k_i\left(1\,+\,2\,P_k(\eta,{\bf p})\,+\,2\,i\frac{{\bf k}\cdot{\bf p}}{\omega_k^2}\delta\theta_k(\eta,{\bf p})\right)+
i\,p_i\,\delta\theta_k(\eta,{\bf p})
\right]\\
\langle T^i_{\;j}(\eta,{\bf p})\rangle\,&=&\,-\frac{1}{(2\pi)^D}\frac{1}{a^{D+1}}\int \text{d}^D {\bf k}\left(\frac{1}{2}\,+\,n_k^T\right)\left[\frac{k_i\,k_j}{\omega_k}\left(1\,+\,2\,\Psi({\bf p})\,+\,2\,P_{k}(\eta,{\bf p})\vphantom{\frac{1}{1}}\right)\,+\,i\,\frac{k_i\,p_j+k_j\,p_i}{\omega_k}\,\delta\theta_k(\eta,{\bf p})\right]\\
\langle T^\mu_{\;\mu}(\eta,{\bf p})\rangle\,&=&\,\frac{1}{(2\pi)^D}\frac{1}{a^{D+1}}\int \text{d}^D {\bf k}\,\left(\frac{1}{2}\,+\,n_k^T\right)\left[\frac{m^2}{\omega_k}\left(1\,+\,2\,P_k(\eta,{\bf p})\vphantom{\frac{1}{1}}\right)\right]\,.
\label{Tmunufull}
\end{eqnarray}
Let us divide the energy-momentum tensor, in the same way as for the potential case, in a vacuum contribution, which does not depend on the number of particles per mode $n_k^T$, and a thermal contribution.
\begin{eqnarray}
\langle T^\mu_{\;\nu}(\eta,{\bf p})\rangle\,=\,\langle T^\mu_{\;\nu}(\eta,{\bf p})\rangle_{\text{vac}}\,+\,\langle T^\mu_{\;\nu}(\eta,{\bf p})\rangle_{T}\,\,.
\end{eqnarray}
each one having a homogeneous and an inhomogeneous part. It can be shown \cite{Higgs} that the energy-momentum tensor of the vaccum is given by $\langle T^\mu_{\;\nu}(\eta,{\bf p})\rangle_{\text{vac}}=\rho_{\text{vac}}\,\delta^\mu_{\;\nu}$, where the energy density $\rho_{\text{vac}}$ and pressure $p_{\text{vac}}$ are given in the $\overline{\text{MS}}$ renormalization scheme with $D=3+\epsilon$ by
\begin{eqnarray}
\rho_{\text{vac}}\,=-\,p_{\text{vac}}\,=\,\frac{m^4}{64\,\pi^2}\,\left[\log\left(\frac{m^2}{\mu^2}\right)\,-\,\frac{3}{2}\right].
\end{eqnarray}
This implies that the inhomogeneous part of the vacuum contribution is zero when dimensional regularization is used, therefore metric perturbations do not contribute to the leading adiabatic order. In this paper, we compute the homogeneous and inhomogeneous parts of $\langle T^\mu_{\;\nu}(\eta,{\bf p})\rangle_{T}$.


\section{Static space-times}
\label{sec:static}

Although the expressions for the perturbed solutions given in Appendix \ref{appsol} are valid for general perturbed FRW space-times, in this work we focus on static space-times, i.e.
we will take $a=1$ and $\Phi=\Phi({\bf x})$, $\Psi=\Psi({\bf x})$.  The general case is of great interest for cosmological scenarios, nevertheless the time dependence of the scale factor increases the complexity of the computations, making extremely difficult to obtain analytical expressions. In addition, in order to define a thermodynamic temperature, there must be a timelike Killing vector field, namely the space-time must be static or stationary. 

In order to compute $V_T(\hat\phi)$ and the energy-momentum tensor $\langle T^\mu_{\;\nu}(\eta,{\bf p})\rangle_{T}$ thermal contributions, our first step will be to expand the $P_k(\eta,{\bf p})$ and $\delta\theta_k(\eta,{\bf p})$ functions in powers of $p\eta$ [Appendix \ref{appexp}]. These expansions, allows us to find a common structure of the integrals involved.

\subsection{Effective potential}
Taking into account \eqref{pserie} and \eqref{effpotT}, it is clear that we have to deal with the following kind of integrals
\begin{eqnarray}
\frac{1}{2}\int_0^{m^2(\hat\phi)}\text{d}m^2\int_0^{\infty}\text{d}k\,k^{D-1}\,\frac{1}{e^{\omega_k/T}-1}\,\frac{1}{\omega_k}\,\left(\frac{k}{\omega_k}\right)^{2\alpha}\,\left(\frac{m}{\omega_k}\right)^{2n}\hspace{0.5cm}\alpha=0,1,2,...\hspace{0.5cm}n=0,1,2
\end{eqnarray}
to compute the finite temperature correction to the effective potential.

It is convenient to use the dimensionless variables $u=\omega_k/T$ and $x=m/T$ instead of $k$ and $m$ respectively. In terms of these new variables the integral reduces to (extracting a global factor $T^{D+1}$)
\begin{eqnarray}
I^X_{\alpha,n}\,&\equiv&\,\int_0^{X}\,\text{d}x\int_x^{\infty}\text{d}u\,\frac{1}{e^{u}-1}\frac{x^{1+2n}}{u^{2\alpha+2n}}\,\left(u^2-x^2\right)^{D/2+\alpha-1}
\end{eqnarray}
where $X\equiv m(\hat\phi)/T$.

It is also useful to interchange the order of integration of this integral and divide it in the following way
\begin{eqnarray}
I^X_{\alpha,n}\,=\,\left(\int_0^{X}\text{d}u\int_0^{u}\text{d}x\,+\,\int_X^{\infty}\text{d}u\int_0^{X}\text{d}x\right)\left(\frac{1}{e^{u}-1}\frac{x^{1+2n}}{u^{2\alpha+2n}}\,\left(u^2-x^2\right)^{D/2+\alpha-1}\right)
\label{div}
\end{eqnarray}
where the first part takes into account the contribution from modes with energies below the mass of the field while the second part includes the contribution from modes with energies above the mass of the field.

\subsection{Energy-momentum tensor}

To compute the energy-momentum tensor, the following integrals appear
\begin{eqnarray}
\int_0^{\infty}\text{d}k\,k^{D-1}\,\frac{1}{e^{\omega_k/T}-1}\,\omega_k\,\left(\frac{k}{\omega_k}\right)^{2\alpha}\,\left(\frac{m}{\omega_k}\right)^{2n}\hspace{0.5cm}\alpha=0,1,2,...\hspace{0.5cm}n=0,1,2
\end{eqnarray}
Using the same dimensionless variables $u$ and $x$ we get (also extracting a global factor $T^{D+1}$)
\begin{eqnarray}
\label{intTmunu}
J^X_{\alpha,n}\,&\equiv&\,\int_X^{\infty}\text{d}u\,\frac{1}{e^{u}-1}\frac{X^{2n}}{u^{2\alpha+2n-2}}\,\left(u^2-X^2\right)^{D/2+\alpha-1}
\end{eqnarray}
Only modes with energies above the mass of the field contribute to the energy-momentum tensor.

In the following we compute the integrals $I^X_{\alpha,n}$ \eqref{div} and $J^{X}_{\alpha,n}$ \eqref{intTmunu} in the non-relativistic and the ultra-relativistic limits.


\section{Non-relativistic limit}
\label{sec:low}

\subsection{Effective potential}

In the non-relativistic limit $m(\hat\phi)/T\rightarrow \infty$ (or $X\rightarrow \infty$), the contribution from modes with energies above the mass of the field is exponentially damped because of the Bose-Einstein factor, hence the leading contribution in the non-relativistic limit is given by the first part of \eqref{div} when taking $X=\infty$
\begin{eqnarray}
I^{\infty}_{\alpha,n}\,&=&\,\int_0^{\infty}\,\text{d}u\int_0^{u}\text{d}x\,\frac{1}{e^{u}-1}\frac{x^{1+2n}}{u^{2\alpha+2n}}\,\left(u^2-x^2\right)^{D/2+\alpha-1}\,\nonumber\\
&=&\,\frac{\Gamma(D/2+\alpha)\,n!}{2\,\Gamma(D/2+\alpha+n+1)}\,D!\,\zeta(D+1)
\label{tinf}
\end{eqnarray}
where $\zeta(x)$ is the Riemann Zeta function.

Therefore, using expression \eqref{effpotT} together with the result for the integral \eqref{tinf} and the expansion of $\hat P_k(\eta,{\bf p})$ \eqref{pserie}, we obtain (assuming $D=3$) for the leading contributions after resummation of the series
in $p\eta$
\begin{eqnarray}
V^{h}_{T(L)}(\hat\phi)\,&=&\,\frac{\pi^2}{90}T^4\\
V^{\Phi}_{T(L)}(\hat\phi)\,&=&\,\frac{\pi^2}{90}T^4\,\Phi({\bf p})\,\times\,4\left[3\frac{\sin(p\,\eta)}{(p\,\eta)^3}-3\frac{\cos(p\,\eta)}{(p\,\eta)^2}-1\right]
\label{resultstinfT}\\
V^{\Psi}_{T(L)}(\hat\phi)\,&=&\,\frac{\pi^2}{90}T^4\,\Psi({\bf p})\,\times\,12\,\left[\left(\frac{6}{(p\,\eta)^4}-\frac{1}{(p\,\eta)^2}\right)\cos(p\,\eta)+\left(\frac{3}{(p\,\eta)^3}-\frac{6}{(p\,\eta)^5}\right)\sin(p\,\eta)\right].
\label{resultstinfS}
\end{eqnarray}
Note that there is no dependence on the field (which may appear through mass terms). Therefore these expressions do not affect the field dynamics and they can be neglected.
On the other hand, even though we are considering static backgrounds, there is an explicit time dependence of the result. This can be traced back to the particular mode choice 
in (\ref{Pk}) and (\ref{thetak}). In particular, taking the $\eta\rightarrow \infty$
limit, which corresponds to setting initial conditions for the modes in the remote past, we recover static results for the effective potential.

In the static limit $\eta\rightarrow\infty$, the following expression is obtained
\begin{eqnarray}
V_{T(L)}(\hat\phi)\,=\,\frac{\pi^2}{90}\,T^4\,\left(1\,-\,4\,\Phi({\bf p})\right)\,.
\end{eqnarray}
It can be shown that the leading inhomogeneous effect in the static limit only depends on the $\Phi$ potential and in fact it   can be obtained from the homogeneous result replacing the temperature by the local Tolman temperature \cite{Tolman}
\begin{eqnarray}
T_{\text{Tolman}}\,=\,\frac{T}{\sqrt{g_{00}}}\,\simeq\,T\left(1-\,\Phi({\bf p})\right)\,.
\label{Tolman}
\end{eqnarray}
Notice however that in the results for finite time given 
in \eqref{resultstinfT} and \eqref{resultstinfS}, the explicit time dependence of the effective potential prevents the 
introduction of a Tolman temperature. 

The next-to-leading correction, $V^{(NL)}_T$, including terms $\Od(T/m(\hat\phi))$, can be obtained by applying a modified version of the Laplace's method to the following integral\footnote{The symbol $\simeq$ stands for an approximation in the Taylor sense, while $\sim$ stands for an asymptotic approximation, namely the quotient between both results equals $1$ in the appropriate limit.}
\begin{eqnarray}
I^{X}_{\alpha,n}\,-\,I^{\infty}_{\alpha,n}\,&\simeq&\,-\int_X^{\infty}\text{d}u\int_X^{u}\text{d}x\,e^{-u}\,\frac{x^{1+2n}}{u^{2\alpha+2n}}\,\left(u^2-x^2\right)^{D/2+\alpha-1}\nonumber\\&=&\,\int_X^{\infty}\text{d}u\,\frac{1}{2}\,e^{-u}\,u^D\,\left[B_{X^2/u^2}\left(1+n,D/2+\alpha\right)\,-\,\frac{\Gamma(1+n)\,\Gamma(D/2+\alpha)}{\Gamma(D/2+\alpha+n+1)}\right]
\end{eqnarray}
where we have replaced the Bose-Einstein factor by the Boltzmann factor. $B_z(a,b)$ is the incomplete Beta function. When $u/X\gg 1$, the integrand is exponentially damped as $e^{-u/X}$. Then, we Taylor expand the expression inside the brackets around $X^2/u^2=1$ to obtain 
\begin{eqnarray}
I^{X}_{\alpha,n}\,-\,I^{\infty}_{\alpha,n}\,\,&{\sim}&\,-\int_X^{\infty}\text{d}u\,\frac{1}{D+2\alpha}\,e^{-u}\,u^D\,\left(1-\frac{X^2}{u^2}\right)^{D/2+\alpha}\nonumber\\&=&\,-\int_X^{\infty}\text{d}u\,\frac{1}{D+2\alpha}\,\exp\left[-u\,+\,D\,\log(u)\,+\,\left(\frac{D}{2}+\alpha\right)\,\log\left(1-\frac{X^2}{u^2}\right)\right]\,.
\end{eqnarray}
The expression inside the exponential has a maximum at $u\sim X$ when $X\rightarrow\infty$.\footnote{Here we are dropping a term linear in $\alpha$ in the expression for the maximum. This means that we cannot allow $\alpha\rightarrow \infty$. Since $\alpha$ is related with the order of the expansion in $p\,\eta$,  
the results are only valid if the series appearing in \eqref{pserie} is truncated at some order such that $\alpha \ll X$.  Although it could be done for arbitrary $\alpha$, it would not be very useful if the expression cannot be ressummed. Nevertheless, it will be shown that the $l$-term is supressed by a factor $1/X^l$, thus only the first terms are relevant in this limit ($X\rightarrow\infty$).} Taylor expanding the argument of the exponential around $u=X$ up to order $O(u)$ [including the logarithmic divergence] the integration in $u$ can be performed to get the following result
\begin{eqnarray}
I^{X}_{\alpha,n}\,-\,I^{\infty}_{\alpha,n}\,\,&{\sim}&\,-2^{3(D/2+\alpha)+1}\,\Gamma(D/2+\alpha)\,\,e^{-X}\,\frac{X^{D+1}}{(4\,X-D+6\alpha)^{D/2+\alpha+1}}\,\nonumber\\&\sim &\,-2^{D/2+\alpha-1}\,\Gamma(D/2+\alpha)\,\,e^{-X}\,X^{D/2-\alpha}\,
\label{finalT0}
\end{eqnarray}
which does not depend on $n$. Because of the factor $X^{D/2-\alpha}$ in the last expression, the expansion in $p\eta$ mixes with the expansion in $X(=m(\hat{\phi})/T)$.

\begin{figure}
    \begin{center}
        {\includegraphics[width=0.48\textwidth]{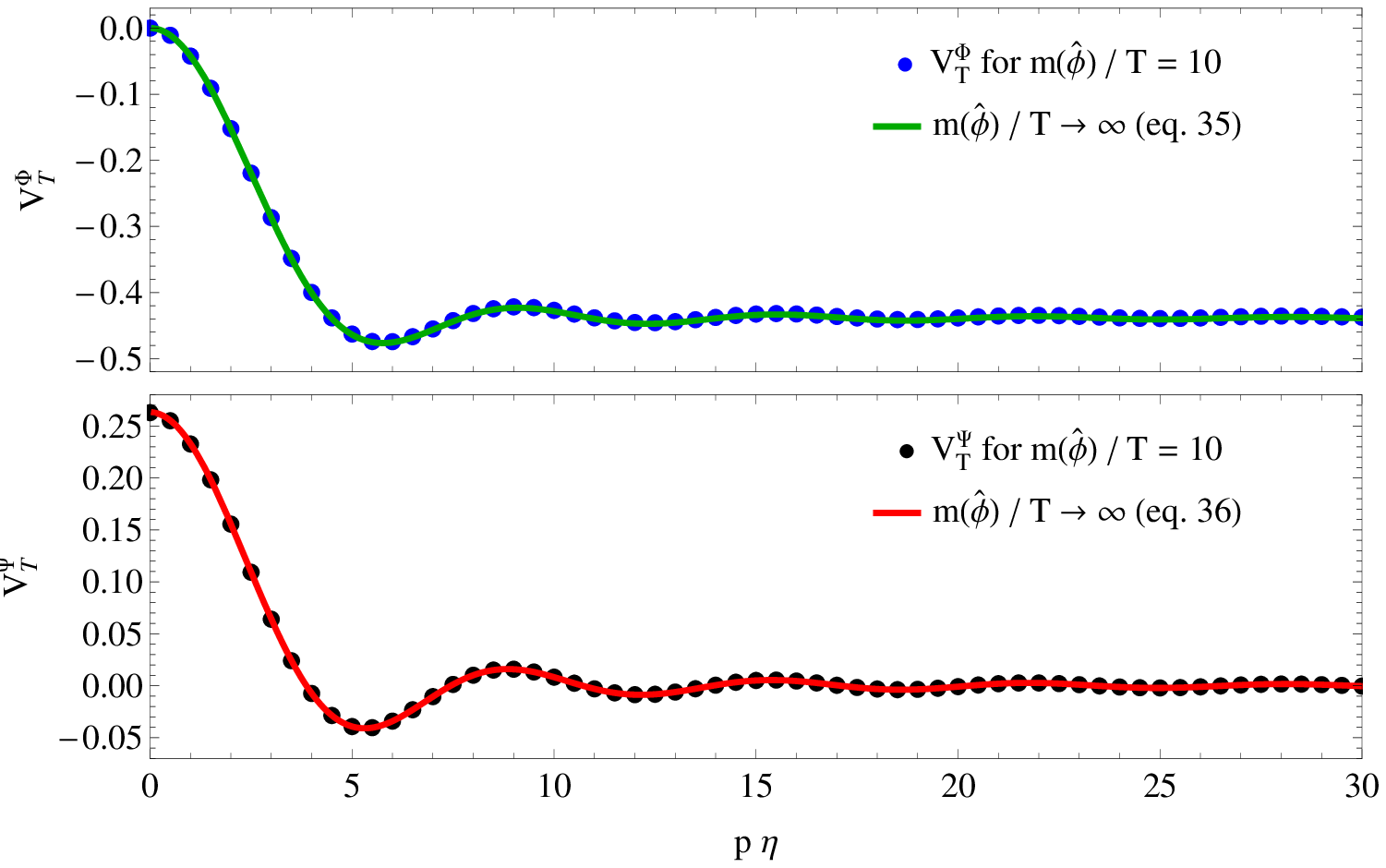}\hspace{0.5cm}\includegraphics[width=0.48\textwidth]{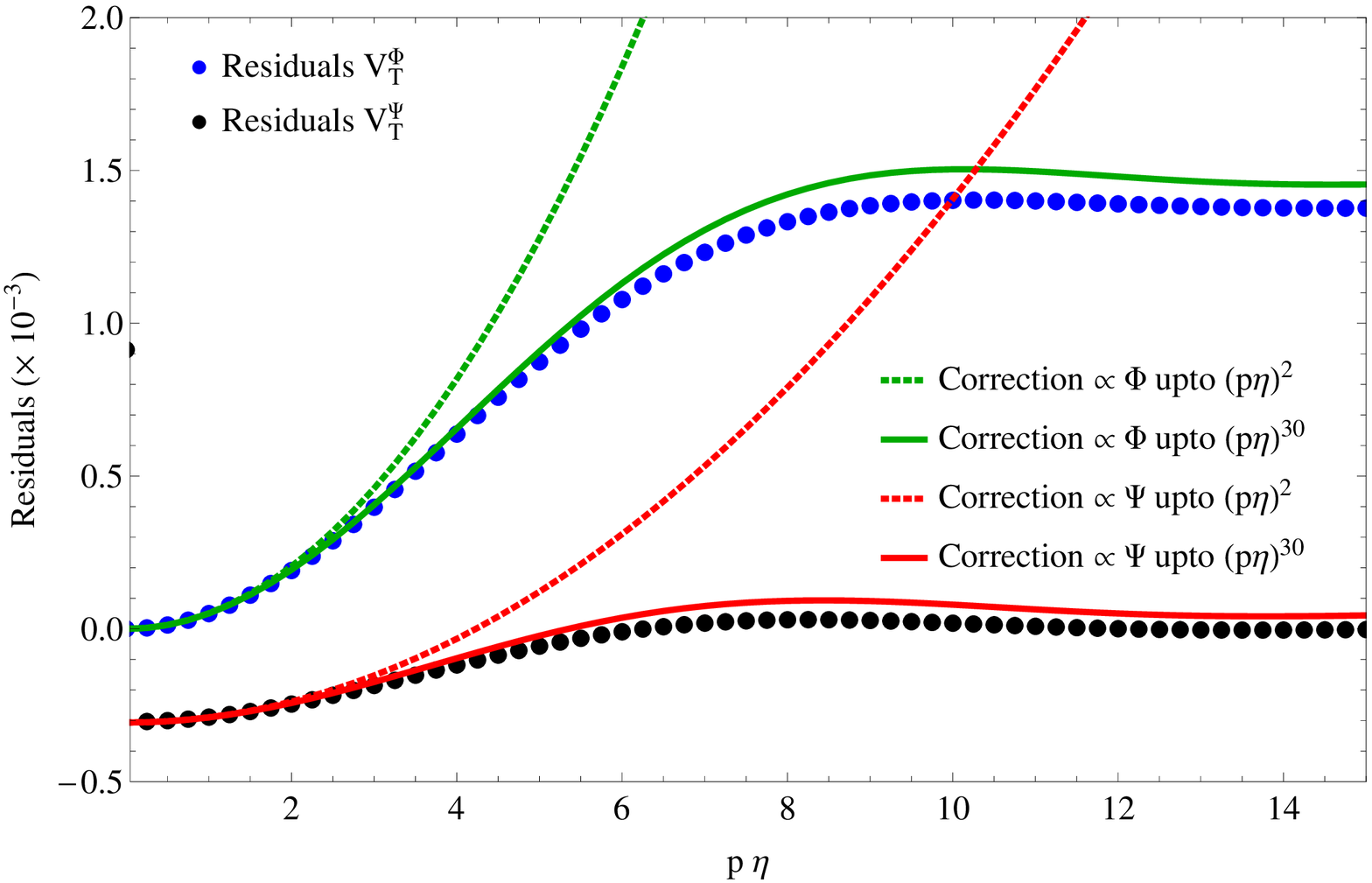}}
    \caption {\footnotesize Left panel: Points correspond to the numerical value of the thermal contributions to the potential proportional to $\Phi$ and $\Psi$ taking $m(\hat\phi)/T=10$, whereas the solid lines represent the leading approximations \eqref{resultstinfT} and \eqref{resultstinfS}. Right panel: Points are the difference between the numerical values of the
    potential  and the approximations \eqref{resultstinfT} (blue points) and \eqref{resultstinfS} (black points)  for $m(\hat\phi)/T=10$. The next-to-leading correction for $m(\hat\phi)/T=10$ up to $(p\eta)^2$ (dashed lines) and $(p\eta)^{30}$ (solid lines) is shown for comparison.
    }
    \label{fig:Vinfinity}
    \end{center}
\end{figure}

Finally, the next to leading contribution to the potential for $p\eta\ll 1$ is given by
\begin{eqnarray}
V_{T(NL)}(\hat \phi)=\,-\,\frac{T^4}{2\sqrt{2}\,\pi^{3/2}}\,e^{-m(\hat\phi)/T}\,\left(\frac{m(\hat\phi)}{T}\right)^{3/2}\left(1\,+\,3\,\Psi({\bf p})\,-\,\frac{(p\eta)^2}{2}\,\Phi({\bf p})\,\right).
\label{correctionT0}
\end{eqnarray}
A better approximation for smaller values of $X$ is obtained if we do not drop $\alpha$ in the denominator in \eqref{finalT0}. This improved approximation is shown in Figure \ref{fig:Vinfinity} (right panel). It is important to note that each order in $(p\eta)$ is suppressed by a factor $(T/m(\hat \phi))$ with respect to the previous order, because of the mixing discussed above. For instance, the correction proportional to $\Psi$ does not depend on $(p\eta)$ to leading order in $(T/m(\hat \phi))$ [see eq. \ref{correctionT0}], then the dependence on $(p\eta)^2$ proportional to $\Psi$ is suppressed by a factor $(T/m(\hat \phi))$ with respect to the $(p\eta)^2$ correction proportional to $\Phi$, as shown in Figure \ref{fig:Vinfinity} (right panel).

Because of the mixing between the expansion in $X(=m(\hat{\phi})/T)$ and $p\eta$ we cannot obtain a result valid for arbitrary scales $p$ and times $\eta$. However, it is possible to obtain the static result by taking the limit $\eta\rightarrow\infty$ directly on \eqref{effpotT}. According to this procedure, we get
\begin{eqnarray}
V_{T(NL)}(\hat \phi)=\,-\,\frac{T^4}{2\sqrt{2}\,\pi^{3/2}}\,e^{-m(\hat\phi)/T}\,\left(\frac{m(\hat\phi)}{T}\right)^{3/2}\left(1\,-\,\frac{35}{8}\,\Phi({\bf p})\,-\,\left(\frac{m(\hat\phi)}{T}\right)\Phi({\bf p})\right).
\label{NLNR}
\end{eqnarray}

As can be checked in a straightforward way from \eqref{NLNR}, also for the next to leading contribution in the static limit, the inhomogeneous correction can be obtained from the homogeneous result by replacing the temperature with the Tolman temperature \eqref{Tolman}.

%

\subsection{Energy-momentum tensor}

The leading order of the energy-momentum tensor is already exponentially damped, since only modes with energies above the mass of the field contribute. We write the integral \eqref{intTmunu} as
\begin{eqnarray}
J^{X}_{\alpha,n}\,{\simeq}\,\int_X^{\infty}\text{d}u\,e^{-u}\,\frac{X^{2n}}{u^{2\alpha+2n-2}}\,\left(u^2-X^2\right)^{D/2+\alpha-1}
\end{eqnarray}
where the Bose-Einstein factor has been replaced by the Boltzmann factor. Applying the Laplace's method again we get
\begin{eqnarray}
\label{JXinf}
J^{X}_{\alpha,n}\,&{\simeq}&\,X^{2n}\,\int_X^{\infty}\text{d}u\,\exp\left[-u\,-2\left(\alpha+n-1\right)\,\log\left(u\right)\,+\,\left(\frac{D}{2}+\alpha-1\right)\log\left(u^2\,-\,X^2\right)\right]\nonumber\\
&\sim & 2^{3(D/2+\alpha)-1	}\,\Gamma(D/2+\alpha)\,\,e^{-X}\,\frac{X^{D+1}}{(4\,X-D+6\alpha+8n-6)^{D/2+\alpha}}\,\nonumber\\&\sim &\,2^{D/2+\alpha-1}\,\Gamma(D/2+\alpha)\,\,e^{-X}\,X^{D/2-\alpha+1}\,
\end{eqnarray}
Then, taking into account the expressions given in Section \ref{sec:back} and the result \eqref{JXinf}, the energy-momentum tensor for $p\eta\ll1$ is given by
\begin{eqnarray}
\rho_T\,\equiv\,\langle T^0_{\;0}(\eta,{\bf p})\rangle_{T}\,&{\sim}&\,\frac{T^4}{2\sqrt{2}\,\pi^{3/2}}\,e^{-m(\hat\phi)/T}\,\left(\frac{m(\hat\phi)}{T}\right)^{5/2}\left(1\,+\,3\Psi({\bf p})\,-\,\frac{(p\eta)^2}{2}\,\Phi({\bf p})\right)\\
-p_T\,\equiv\,\langle T^i_{\;i}(\eta,{\bf p})\rangle_{T}\,&{\sim}&\,-\frac{T^4}{2\sqrt{2}\,\pi^{3/2}}\,e^{-m(\hat\phi)/T}\,\left(\frac{m(\hat\phi)}{T}\right)^{3/2}\left(1\,+\,5\Psi({\bf p})\,-\,\frac{5(p\eta)^2}{6}\,\Phi({\bf p})\right)\,\\
\langle T^i_{\;0}(\eta,{\bf p})\rangle_{T}\,&{\sim}&\,-\frac{T^4}{2\sqrt{2}\,\pi^{3/2}}\,e^{-m(\hat\phi)/T}\,\left(\frac{m(\hat\phi)}{T}\right)^{5/2}\,(ip_i)\,\eta\,\Phi({\bf p})\\\
\langle T^i_{\;j}(\eta,{\bf p})\rangle_{T}\,&{\sim}&\,-\frac{T^4}{2\sqrt{2}\,\pi^{3/2}}\,e^{-m(\hat\phi)/T}\,\left(\frac{m(\hat\phi)}{T}\right)^{3/2}(i^2p_ip_j)\,\eta^2\,\Phi({\bf p})\hspace{1cm}i\neq j
\end{eqnarray}
where $\rho_T$ and $p_T$ are the energy density and pressure produced by the thermal corrections. We have only retained the leading order in $m(\hat \phi)/T$. 
Further corrections $\Od((p\eta)^{2l})$ are suppressed by a factor $(m(\hat \phi)/T)^l$.

In the non-relativistic case, it is not possible to take the static limit in the final expressions since we only have the results for $p\eta\ll1$ as discussed before. However, the static expression can be obtained by taking the static limit in the original expressions \eqref{Tmunufull}
\begin{eqnarray}
\rho_T\,\equiv\,\langle T^0_{\;0}(\eta,{\bf p})\rangle_{T}\,&{\sim}&\,\,\frac{T^4}{2\sqrt{2}\,\pi^{3/2}}\,e^{-m(\hat\phi)/T}\,\left(\frac{m(\hat\phi)}{T}\right)^{5/2}\left(1\,-\,\frac{39}{8}\,\Phi({\bf p})\,-\,\left(\frac{m(\hat\phi)}{T}\right)\Phi({\bf p})\right)\\
-p_T\,\equiv\,\langle T^i_{\;i}(\eta,{\bf p})\rangle_{T}\,&{\sim}&\,-\,\frac{T^4}{2\sqrt{2}\,\pi^{3/2}}\,e^{-m(\hat\phi)/T}\,\left(\frac{m(\hat\phi)}{T}\right)^{3/2}\left(1\,-\,\frac{35}{8}\,\Phi({\bf p})\,-\,\left(\frac{m(\hat\phi)}{T}\right)\Phi({\bf p})\right).
\end{eqnarray}
Once again, in the static limit, the inhomogeneous corrections, depending only on 
the $\Phi$ potential and can be obtained from the homogenous 
one by introducing the Tolman temperature.

%
%
%
\section{Ultra-relativistic limit}
\label{sec:high}

\subsection{Effective potential}

In the ultra-relativistic limit, $m(\hat\phi)/T\rightarrow 0$ (or $X\rightarrow 0$), the dominant contribution comes from modes with energies higher than the mass of the field. Therefore, the second part of \eqref{div} gives
\begin{eqnarray}
I^{X}_{\alpha,n}\,&{\simeq}&\,\int_X^{\infty}\text{d}u\int_0^{X}\text{d}x\,\frac{1}{e^{u}-1}\frac{x^{1+2n}}{u^{2\alpha+2n}}\,\left(u^2-x^2\right)^{D/2+\alpha-1}\nonumber\\&=&\,\int_X^{\infty}\text{d}u\,\frac{1}{2}\,\frac{u^D}{e^{u}-1}\,B_{X^2/u^2}(1+n,\,D/2+\alpha)\nonumber\\&{\simeq}&\int_X^{\infty}\text{d}u\,\frac{1}{2}\,\frac{u^{D-2n-2}}{e^{u}-1}\,\frac{X^{2+2n}}{1+n}
\label{IPot}
\end{eqnarray}
where we have expanded the incomplete Beta function $B_z(a,b)$  for $X\ll 1$ in the last line. The leading contribution comes from $n=0$. Replacing the lower limit of integration by 0 we get in that limit
\begin{eqnarray}
I^{X}_{\alpha,0}\,&{\simeq}&\,\int_0^{\infty}\text{d}u\,\frac{1}{2}\,\frac{u^{D-2}}{e^{u}-1}\,X^{2}\nonumber\\
&=&\frac{1}{2}\,\Gamma(D-1)\,\text{Li}_{D-1}(1)\,X^2
\label{t0}
\end{eqnarray}
where $\text{Li}_n(z)$ is the polylogarithm function.

Therefore, from \eqref{div} and using the expansion of $\hat P_k(\eta,{\bf p})$ in \eqref{pserie} and the result \eqref{t0}, we can resum this contribution to get
the leading contribution
\begin{eqnarray}
V^{h}_{T(L)}(\hat\phi)\,&=&\,\frac{T^4}{24}\left(\frac{m(\hat\phi)}{T}\right)^2\\
V^{\Phi}_{T(L)}(\hat\phi)\,&=&\,\frac{T^4}{12}\left(\frac{m(\hat\phi)}{T}\right)^2\,\Phi({\bf p})\,\times\left(\frac{\sin(p\,\eta)}{p\,\eta}\,-\,1\right)
\label{resultst0T}\\
V^{\Psi}_{T(L)}(\hat\phi)\,&=&\,\frac{T^4}{12}\left(\frac{m(\hat\phi)}{T}\right)^2\,\Psi({\bf p})\,\times\left(\frac{\sin(p\,\eta)}{p\,\eta}\right)
\label{resultst0S}
\end{eqnarray}
%

%
%

The explicit time dependence of the general results obtained in a static metric can be traced back to the initial conditions of the modes. Taking the limit $\eta\rightarrow\infty$ 
in (\ref{resultst0T}) and (\ref{resultst0S}), the initial conditions are washed out and the remaining correction in Fourier space is
\begin{eqnarray}
V_{T(L)}(\hat\phi)\,&=&\,\frac{T^4}{24}\left(\frac{m(\hat\phi)}{T}\right)^2\left(1\,-\,2\,\Phi({\bf p})\vphantom{\frac{1}{1}}\right)\,.
\label{staticp}
\end{eqnarray}
In this case we can also obtain the inhomogeneous result by replacing the temperature by the local Tolman temperature \eqref{Tolman} in the homogeneous result.

To get the real space result in the static limit, one has to compute the Fourier transform of the complete expression and then take the static limit, $\eta\rightarrow\infty$. Following this procedure, it is possible to get the real space result for arbitrary perturbation (see Appendix \ref{apppoles}) which reads
\begin{eqnarray}
V_{T(L)}(\hat\phi)\,&=&\,\frac{T^4}{24}\left(\frac{m(\hat\phi)}{T}\right)^2\left(1\,-\,2\,\Phi({\bf r})\vphantom{\frac{1}{1}}\right)\,.
\label{staticr}
\end{eqnarray}
Therefore, as expected, the static limit and the Fourier transform commute (compare \eqref{staticp} and \eqref{staticr}). This is a general conclusion for the functions in Fourier space appearing in this paper due to the results of Appendix \ref{apppoles}.

In real space, the corrections due to Newtonian perturbations $\Phi_N({\bf p})$ and $\Psi_N({\bf p})$ given by
\begin{eqnarray}
\Phi_N({\bf p})\,&=&\,\Psi_N({\bf p})=-4\,\pi\frac{GM}{p^2}\\
\Phi_N({\bf r})\,&=&\,\Psi_N({\bf r})=-\frac{GM}{r}\,,
\end{eqnarray}
inside the lightcone ($r<|\eta|$) are
\begin{eqnarray}
V^{\Phi_N}_{T(L)}(\hat\phi)\,&=&\,\frac{T^4}{12}\left(\frac{m(\hat\phi)}{T}\right)^2\,\Phi_N({\bf r})\,\times\left(\frac{r}{|\eta|}-1\right)\\
V^{\Psi_N}_{T(L)}(\hat\phi)\,&=&\,\frac{T^4}{12}\left(\frac{m(\hat\phi)}{T}\right)^2\,\Psi_N({\bf r})\,\times\left(\frac{r}{|\eta|}\,\right)
\label{resultst0real}
\end{eqnarray}
while on and outside the lightcone ($r\geq|\eta|$) are
\begin{eqnarray}
V^{\Phi_N}_{T(L)}(\hat\phi)\,&=&\,0\\
V^{\Psi_N}_{T(L)}(\hat\phi)\,&=&\,\frac{T^4}{12}\left(\frac{m(\hat\phi)}{T}\right)^2\,\Psi_N({\bf r})\,.
\label{resultst0real}
\end{eqnarray}

The next-to-leading order corrections can be obtained by computing the first part of equation \eqref{div} plus next-to-leading terms coming from equation \eqref{IPot} (see Appendix \ref{appnexto}). Finally, after resummation of the  the series expansion \eqref{pserie} we get for $V_{T(NL)}$, (up to $O((m/T)^5)$)
\begin{eqnarray}
V^{h}_{T(NL)}(\hat\phi)\,&{=}&\,T^4\left(\frac{m(\hat\phi)}{T}\right)^3\left[-\frac{1}{12\,\pi}\,+\,\frac{1}{32\,\pi^2}\left(\frac{m(\hat\phi)}{T}\right)\left(\log\left(\frac{T}{M}\right)\,+\,\frac{3}{4}\,-\,\gamma\,+\,\log(4\,\pi)\,
\right)\right]\\
V^{\Phi}_{T(NL)}(\hat\phi)\,&{=}&\,T^4\,\left(\frac{m(\hat\phi)}{T}\right)^3\,\Phi({\bf p})\left[-\frac{1}{12\,\pi}
\,\left(J_0(p\,\eta)\,-\,1\right)\,+\,\frac{1}{32\,\pi^2}\left(\frac{m(\hat\phi)}{T}\right)
\,\left(\cos(p\,\eta)\,-\,1\right)\,+\,\frac{1}{720\,\pi}\left(\frac{m(\hat\phi)}{T}\right)^2
\,p\,\eta\,J_1(p\,\eta)\right]\nonumber\\
\label{resultst1T}\\
V^{\Psi}_{T(NL)}(\hat\phi)\,&{=}&\,T^4\,\left(\frac{m(\hat\phi)}{T}\right)^3\Psi({\bf p})\left[-\frac{1}{12\,\pi}
\,\left(J_0(p\,\eta)\,+\,\frac{J_1(p\,\eta)}{p\,\eta}\right)\,+\,\frac{1}{32\,\pi^2}\left(\frac{m(\hat\phi)}{T}\right)
\,\left(\cos(p\,\eta)\,+\,\frac{2\,\sin{p\,\eta}}{p\,\eta}\right)\,+\,\right.\nonumber\\&&+\,\left.\frac{1}{720\,\pi}\left(\frac{m(\hat\phi)}{T}\right)^2
\,\left(p\,\eta\,J_1(p\,\eta)\,-\,3\,J_0(p\,\eta)\right)\right]
\label{resultst1S}
\end{eqnarray}
where $\gamma$ is Euler's constant and $J_n(x)$  Bessel functions.

%
%

Considering Newtonian perturbations $\Phi_N$ and $\Psi_N$, in real space we get for the region inside the lightcone ($r<|\eta|$)
\begin{eqnarray}
V^{\Phi_N}_{T(NL)}(\hat\phi)\,&{=}&\,T^4\left(\frac{m(\hat\phi)}{T}\right)^3
\,\Phi_N({\bf r})\,\times\left[\frac{1}{12\,\pi}\,-\,\frac{1}{6\,\pi^2}\arcsin\left(\frac{r}{|\eta|}\right)\right]\\
V^{\Psi_N}_{T(NL)}(\hat\phi)\,&{=}&\,T^4\left(\frac{m(\hat\phi)}{T}\right)^3
\,\Psi_N({\bf r})\,\times\left[-\frac{1}{12\,\pi^2}\frac{r}{\eta}\sqrt{1-\frac{r^2}{\eta^2}}\,-\,\frac{1}{4\,\pi^2}\arcsin\left(\frac{r}{|\eta|}\right)\right]\,,
\label{resultst1real}
\end{eqnarray}
and outside and on the lightcone  ($r\geq|\eta|$)
\begin{eqnarray}
V^{\Phi_N}_{T(NL)}(\hat\phi)\,&{=}&\,0\\
V^{\Psi_N}_{T(NL)}(\hat\phi)\,&{=}&\,T^4\left(\frac{m(\hat\phi)}{T}\right)^3
\,\Psi_N({\bf r})\,\times\left[-\frac{1}{8\,\pi}\right]\,.
\label{resultst1real}
\end{eqnarray}
Here for simplicity we have only shown the $\Od(m/T)^3$ contributions.

In the static limit $\eta\rightarrow\infty$ one gets
\begin{eqnarray}
V_{T(NL)}(\hat\phi)\,&{=}&\,-\frac{T^4}{12\,\pi}\left(\frac{m(\hat\phi)}{T}\right)^3\,\left(\vphantom{\frac{1}{1}}1\,-\,\Phi({\bf p})\right)\,,
\label{resultst1p}
\end{eqnarray}
which is also valid in real space replacing $\Phi({\bf p})$ by $\Phi({\bf r})$ (see Appendix \ref{apppoles}). Here again we find that the inhomogeneous result can be obtained by replacing the temperature in the homogeneous contribution by the local Tolman temperature.
\begin{figure}
    \begin{center}
        {\includegraphics[width=0.49\textwidth]{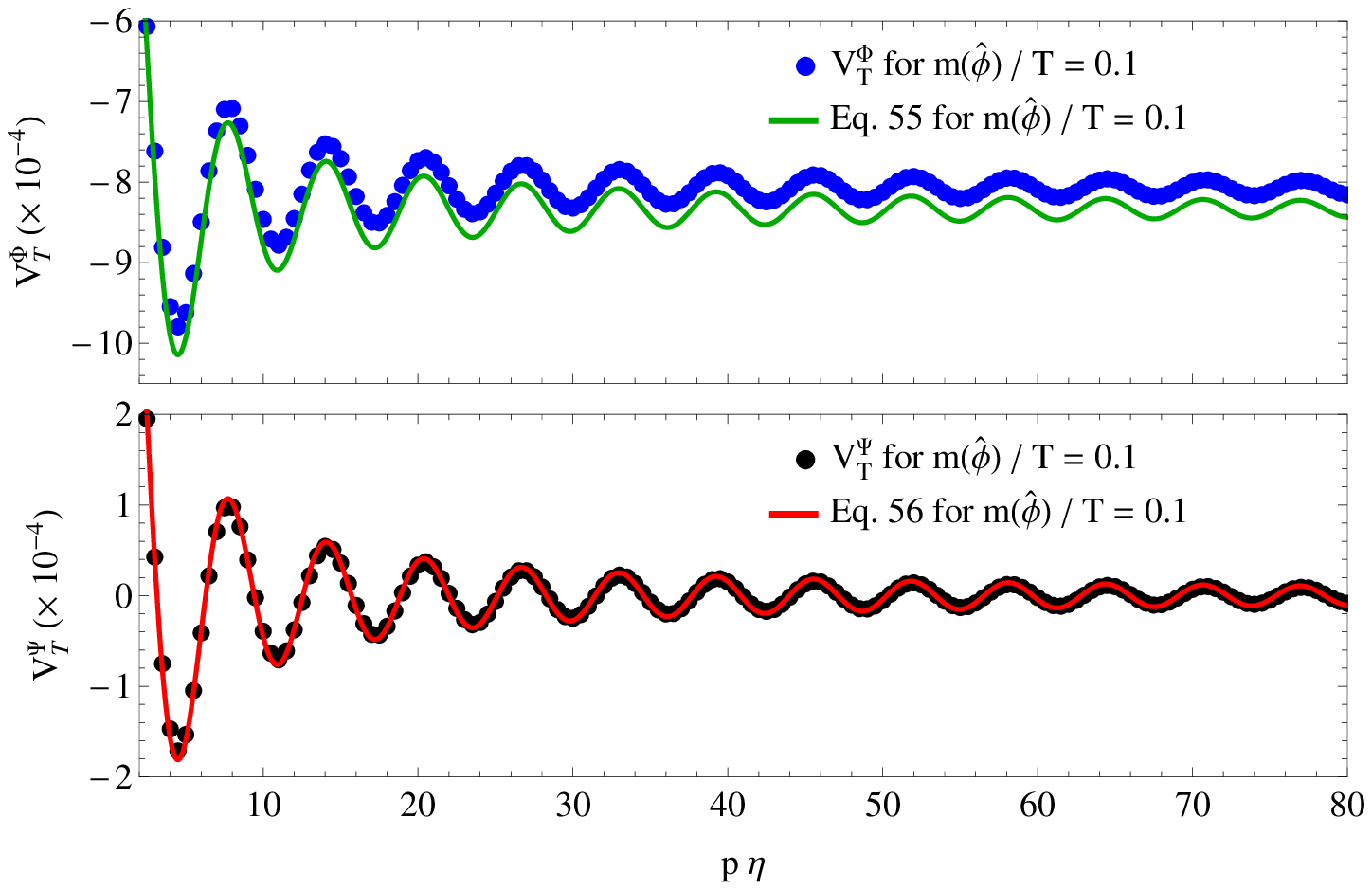}\hspace{0.2cm}\includegraphics[width=0.49\textwidth]{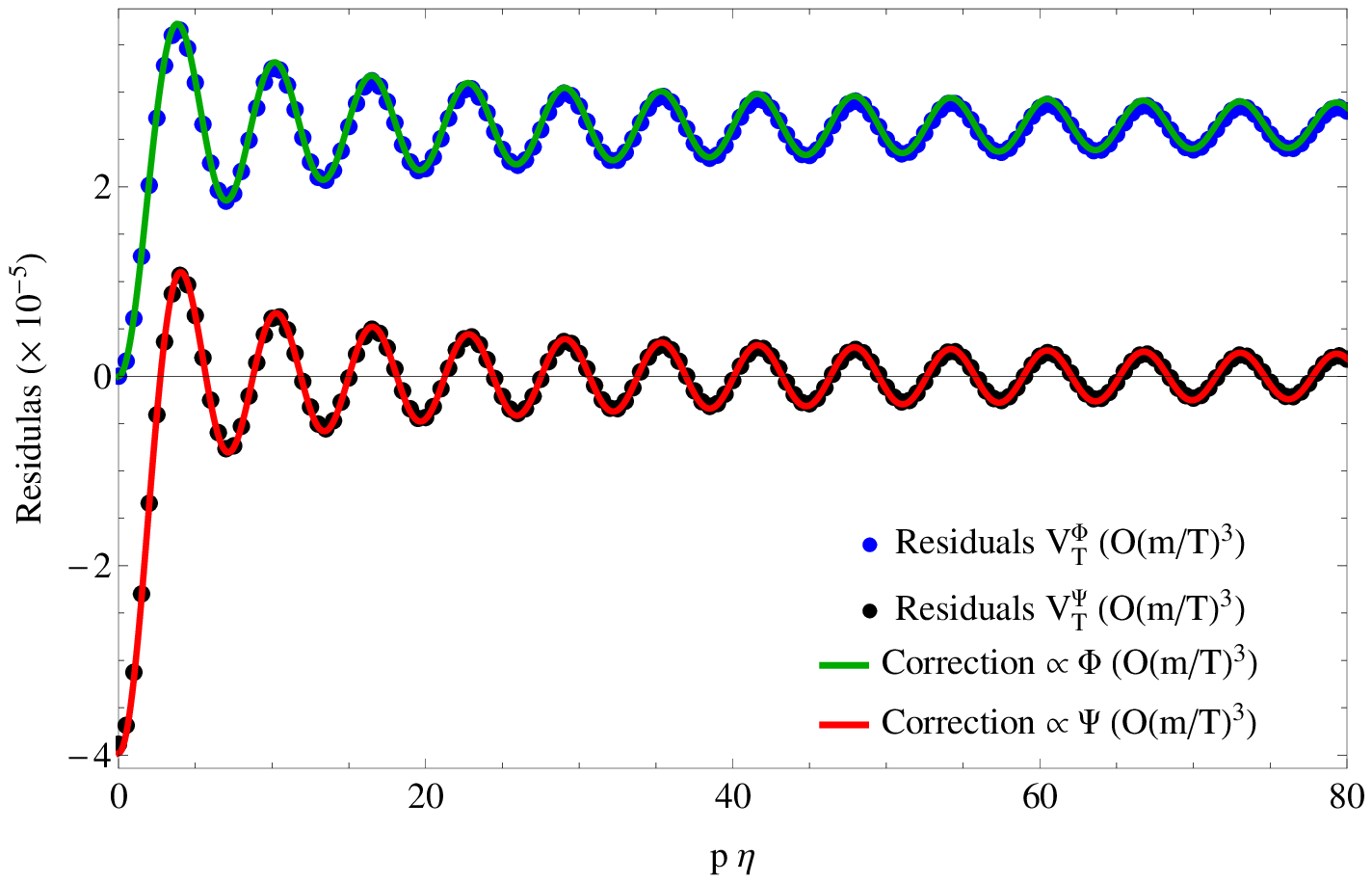}}
             {\includegraphics[width=0.49\textwidth]{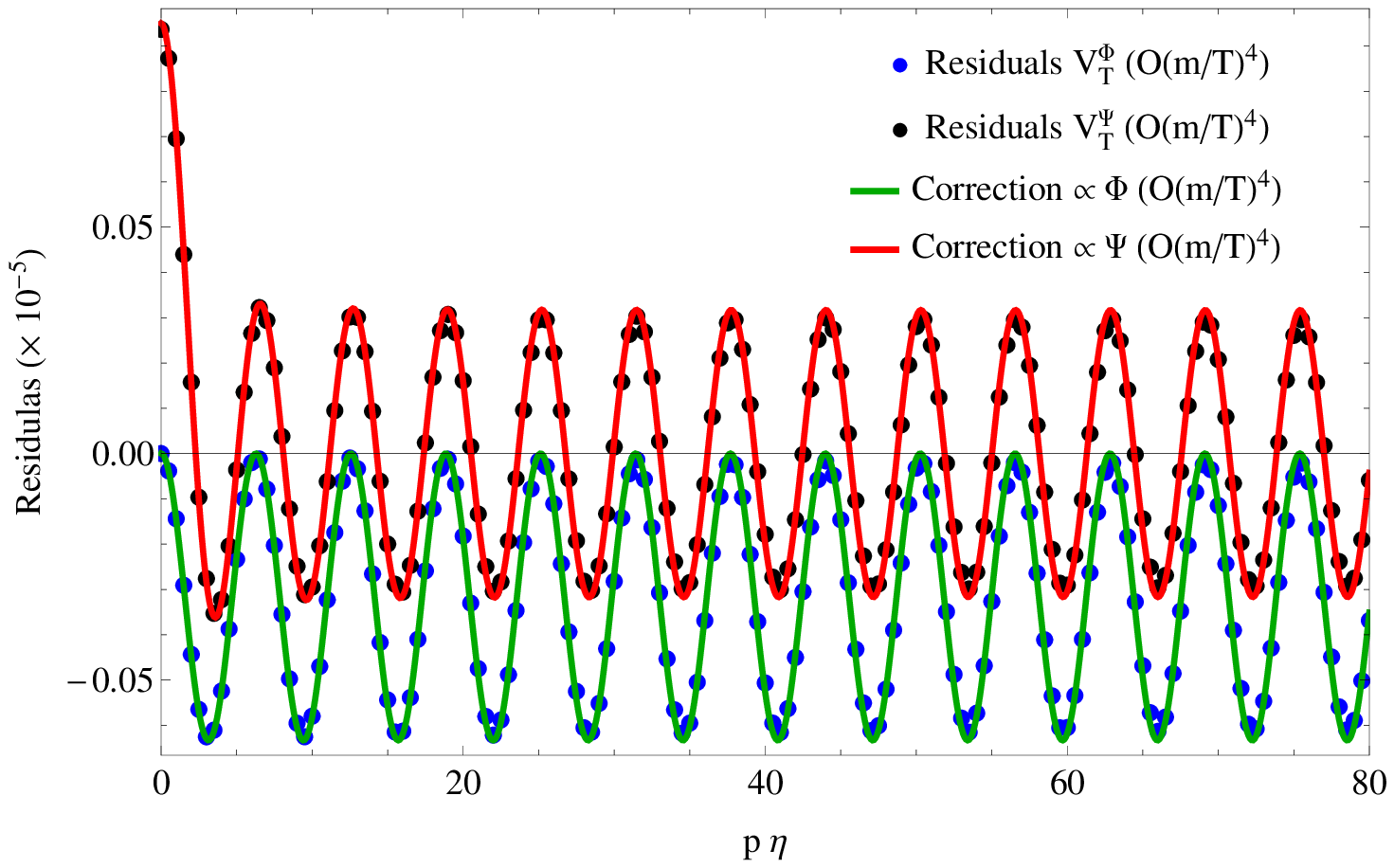}\hspace{0.2
       cm}\includegraphics[width=0.49\textwidth]{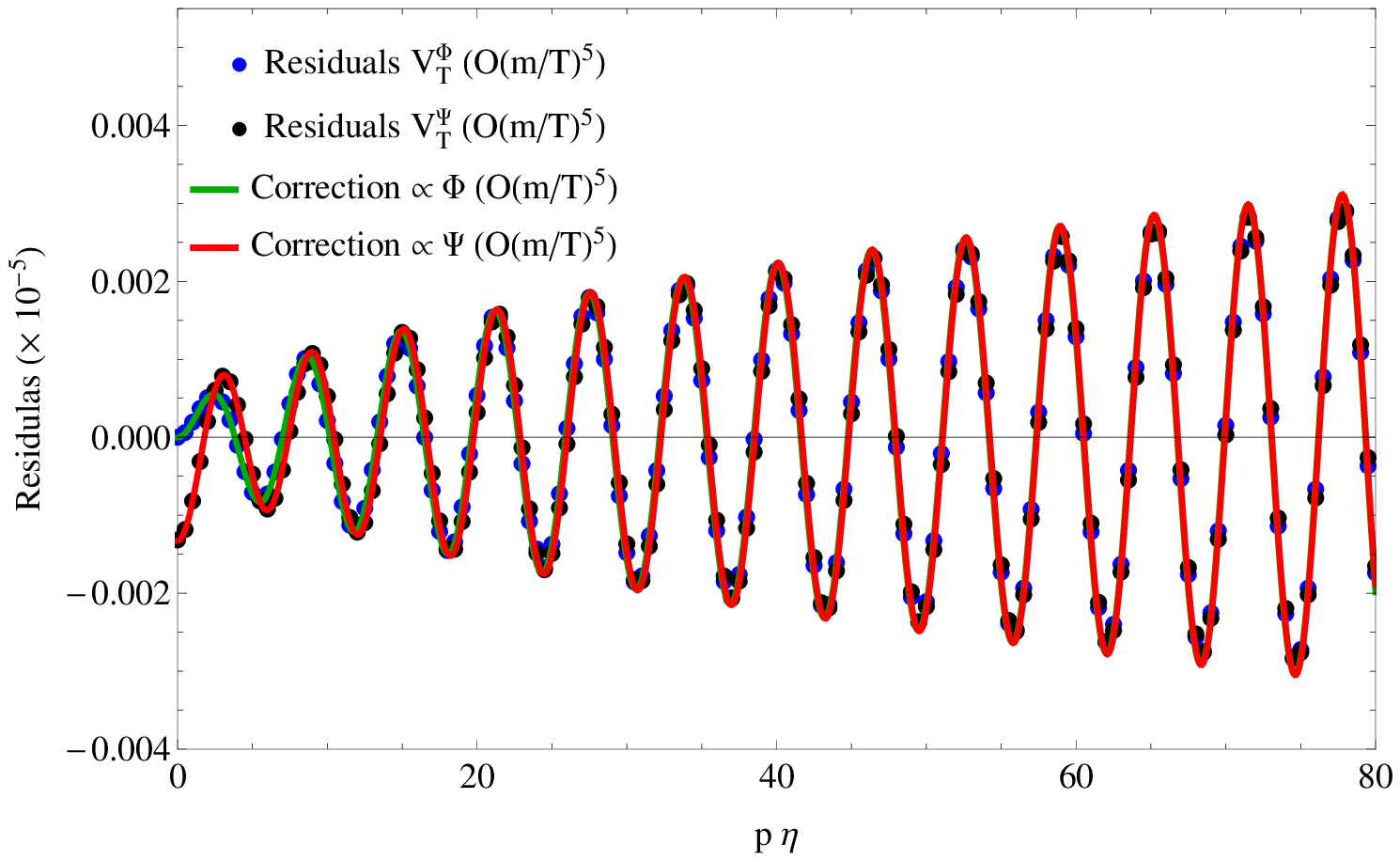}}
    \caption {\footnotesize Left upper panel: Points show the numerical value of the thermal contribution to the potential taking $m(\hat\phi)/T=0.1$ and the continuous line corresponds to the approximations in \eqref{resultst0T} and \eqref{resultst0S}. Right upper panel:  Difference between the
 numerical value of the potential for $m(\hat\phi)/T=0.1$ and the approximations \eqref{resultst0T} (blue points) or  \eqref{resultst0S} (black points). The next-to-leading corrections  ($O((m/T)^3)$) given by \eqref{resultst1T} (green solid line) and \eqref{resultst1S} (red solid line) for $m(\hat\phi)/T=0.1$. Left bottom panel: Difference between the numerical value and the $O((m/T)^3)$ approximation (blue and black points). The $O((m/T)^4)$ correction is plotted as a solid line. Right bottom panel: Difference between the numerical value and the $O((m/T)^4)$ approximation (blue and black points). The $O((m/T)^5)$ correction is plotted as a solid line.
    }
    \label{fig:V0}
    \end{center}
\end{figure}
%

\subsection{Energy-momentum tensor}
The leading contribution is given by the integral \eqref{intTmunu} when $n=0$
\begin{eqnarray}
J^{X}_{\alpha,0}\,&=&\,\int_X^{\infty}\text{d}u\,\frac{1}{e^{u}-1}\frac{1}{u^{2\alpha-2}}\,\left(u^2-X^2\right)^{D/2+\alpha-1}\nonumber\\
&{\simeq}&\,\int_0^{\infty}\text{d}u\,\frac{u^{D-2}}{e^u-1}\left(\vphantom{\frac{1}{1}}u^2\,+\,(1-D/2-\alpha)X^2\right)\nonumber\\
&=&\,\Gamma(D-1)\left[\vphantom{\frac{1}{1}}(D-1)D\,\zeta(D+1)\,+\,(1-D/2-\alpha)\zeta(D-1)\,X^2\right]
\end{eqnarray}
where we have replaced the lower limit of integration by $0$ and expanded the integrand around $X=0$ in the second line. Therefore, we get for the energy-momentum tensor
\begin{eqnarray}
\label{00}
\frac{\rho_T}{T^4}\,=\,\frac{\langle T^0_{\;0}(\eta,{\bf p})\rangle_{T}}{T^4}\,&{\simeq}&\,\frac{\pi^2}{30}\left(1\,-\,4\Phi({\bf p})\,+\,\frac{4\sin(p\eta)}{p\eta}(\Phi({\bf p})+\Psi({\bf p}))\right)\,\\&&-\frac{1}{24}\left(\frac{m(\hat\phi)}{T}\right)^2\left[1\,+\left(2\cos(p\eta)\,+\,\frac{4\sin(p\eta)}{p\eta}\right)\Psi({\bf p})+\left(2\cos(p\eta)\,-\,2\right)\Phi({\bf p})\right]\nonumber\\
\label{ii}
\frac{p_T}{T^4}\,=\,-\frac{\langle T^i_{\;i}(\eta,{\bf p})\rangle_{T}}{T^4}\,&{\simeq}&\,\frac{\pi^2}{90}\left(1\,-\,4\Phi({\bf p})\,+\,\frac{4\sin(p\eta)}{p\eta}(\Phi({\bf p})+\Psi({\bf p}))\right)\,\\&&-\,\frac{1}{24}\left(\frac{m(\hat\phi)}{T}\right)^2\left[1\,+\,\left(\frac{2}{3}\cos(p\eta)\,+\,\frac{8\sin(p\eta)}{3p\eta}\right)\Psi({\bf p})\,+\,\left(\frac{2}{3}\cos(p\eta)\,+\,\frac{4\sin(p\eta)}{3p\eta}\,-\,2\right)\Phi({\bf p})\right]\nonumber\\
\label{i0}
\frac{\langle T^i_{\;0}(\eta,{\bf p})\rangle_{T}}{T^4}\,&{\simeq}&\,\left(i\frac{p_i}{p}\right)\frac{2\pi^2}{15} \left(\frac{\cos(p\eta)}{p\eta}\,-\,\frac{\sin(p\eta)}{(p\eta)^2}\right)(\Phi({\bf p})\,+\,\Psi({\bf p}))\,\\&&+\left(i\frac{p_i}{p}\right)\,\frac{1}{12}\left(\frac{m(\hat\phi)}{T}\right)^2\,\left[\sin(p\eta)\,\Phi({\bf p})\,+\,\left(\sin(p\eta)\,+\,2\frac{\sin(p\eta)}{(p\eta)^2}\,-\,2\frac{\cos(p\eta)}{p\eta}\right)\Psi({\bf p})\right]\nonumber\\
\label{ij}
\frac{\langle T^i_{\;j}(\eta,{\bf p})\rangle_{T}}{T^4}\,&{\simeq}&\,\left(i^2\frac{p_ip_j}{p^2}\right)\frac{2\pi^2}{15}\left(\frac{\sin(p\eta)}{p\eta}+\,3\frac{\cos(p\eta)}{(p\eta)^2}-3\frac{\sin(p\eta)}{(p\eta)^3}\,\right)(\Phi({\bf p})\,+\Psi({\bf p}))\\&&+\left(i^2\frac{p_ip_j}{p^2}\right)\,\frac{1}{12}\left(\frac{m(\hat\phi)}{T}\right)^2\,\times\nonumber\\&&\left[\left(\frac{\sin(p\eta)}{p\eta}\,-\,\cos(p\eta)\right)\Phi({\bf p})\,+\,\left(6\frac{\sin(p\eta)}{(p\eta)^3}\,-\,\frac{\sin(p\eta)}{p\eta}-\,6\frac{\cos(p\eta)}{(p\eta)^2}\,-\,\cos(p\eta)\right)\Psi({\bf p})\right]\hspace{0,5cm}i\neq j\nonumber\\
\frac{\langle T^\mu_{\;\mu}(\eta,{\bf p})\rangle_{T}}{T^4}\,&{\simeq}&\,\frac{1}{12}\left(\frac{m(\hat\phi)}{T}\right)^2\left[1\,+\,\left(\frac{2\sin(p\eta)}{p\eta}\,-\,2\right)\Phi({\bf p})\,+\,\frac{2\sin(p\eta)}{p\eta}\,\Psi({\bf p})\right]
\label{tr}
\end{eqnarray}
which does not correspond to a perfect fluid\footnote{The energy-momentum tensor given by equations \eqref{00}, \eqref{ii}, \eqref{i0} and \eqref{ij} is conserved.}.

%
%

In real space, we have for Newtonian perturbations inside the lightcone ($r<|\eta|$)
\begin{eqnarray}
\frac{\rho_T}{T^4}\,=\,\frac{\langle T^0_{\;0}(\eta,{\bf r})\rangle_{T}}{T^4}\,&\,{\simeq}\,&\,\frac{\pi^2}{30}\left\lbrace 1\,-\,4\left[\left(1-\frac{r}{|\eta|}\right)\Phi_N({\bf r})\,-\,\frac{r}{|\eta|}\Psi_N({\bf r})\right]\right\rbrace\,-\frac{1}{24}\left(\frac{m(\hat\phi)}{T}\right)^2\left[1\,-\,2\,\Phi_N({\bf r})\,+\,4\,\frac{r}{|\eta|}\Psi_N({\bf r})\vphantom{\frac{1}{1}}\right]\nonumber\\\ \\
\frac{p_T}{T^4}\,=\,-\frac{\langle T^i_{\;i}(\eta,{\bf r})\rangle_{T}}{T^4}\,&\,{\simeq}\,&\,\frac{\pi^2}{90}\left\lbrace 1\,-\,4\left[\left(1-\frac{r}{|\eta|}\right)\Phi_N({\bf r})\,-\,\frac{r}{|\eta|}\Psi_N({\bf r})\right]\right\rbrace\,\\&&-\frac{1}{24}\left(\frac{m(\hat\phi)}{T}\right)^2\left[1\,-\,2\,\left(1-\frac{2}{3}\frac{r}{|\eta|}\right)\Phi_N({\bf r})\,+\,\frac{8}{3}\frac{r}{|\eta|}\Psi_N({\bf r}))\vphantom{\frac{1}{1}}\right]\nonumber\\
\ 
\frac{\langle T^i_{\;0}(\eta,{\bf r})\rangle_{T}}{T^4}\,&{\simeq}&\,\frac{\pi^2}{45\,\eta^2}\,\partial_i\left[\vphantom{\frac{1}{1}}r^3\,\left(\Phi_N({\bf r})\,+\,\Psi_N({\bf r})\right)\right]\,-\frac{1}{36\,\eta^2}\left(\frac{m(\hat\phi)}{T}\right)^2\partial_i\left(\vphantom{\frac{1}{1}}r^3\,\Psi_N({\bf r})\right)\\
\ 
\frac{\langle T^i_{\;j}(\eta,{\bf r})\rangle_{T}}{T^4}\,&{\simeq}&\,-\frac{\pi^2}{300\,|\eta|^3}\partial_i\partial_j\left[\vphantom{\frac{1}{1}}r^5\,\left(\Phi_N({\bf r})\,+\,\Psi_N({\bf r})\right)\right]\,+\,\frac{1}{240\,|\eta|^3}\left(\frac{m(\hat\phi)}{T}\right)^2\partial_i\partial_j\left(\vphantom{\frac{1}{1}}r^5\,\Psi_N({\bf r})\right)\hspace{0,5cm} i\neq j\nonumber\\ \ \\
\frac{\langle T^\mu_{\;\mu}(\eta,{\bf r})\rangle_{T}}{T^4}\,&\,{\simeq}\,&\,\frac{1}{12}\left(\frac{m(\hat\phi)}{T}\right)^2\left[1\,-2\,\left(1-\frac{r}{|\eta|}\right)\Phi_N({\bf r})\,+\,2\,\frac{r}{|\eta|}\Psi_N({\bf r})\vphantom{\frac{1}{1}}\right].
\end{eqnarray}
Outside the lightcone ($r>|\eta|$), we get
\begin{eqnarray}
\frac{\rho_T}{T^4}\,=\,\frac{\langle T^0_{\;0}(\eta,{\bf r})\rangle_{T}}{T^4}\,&\,{\simeq}\,&\,\frac{\pi^2}{30}\left(1\,+\,4\,\Psi_N({\bf r})\vphantom{\frac{1}{1}}\right)\,-\frac{1}{24}\left(\frac{m(\hat\phi)}{T}\right)^2\left[1\,+\,6\,\Psi_N({\bf r})\vphantom{\frac{1}{1}}\right]\\
\frac{p_T}{T^4}\,=\,-\frac{\langle T^i_{\;i}(\eta,{\bf r})\rangle_{T}}{T^4}\,&\,{\simeq}\,&\,\frac{\pi^2}{90}\left( 1\,+\,4\,\Psi_N({\bf r})\vphantom{\frac{1}{1}}\right)\,-\frac{1}{24}\left(\frac{m(\hat\phi)}{T}\right)^2\left[1\,+\,\frac{10}{3}\,\Psi_N({\bf r}))\vphantom{\frac{1}{1}}\right]\\
\ 
\frac{\langle T^i_{\;0}(\eta,{\bf r})\rangle_{T}}{T^4}\,&{\simeq}&\,-\frac{2\,\pi^2\,\eta}{45}\,\partial_i\left(\vphantom{\frac{1}{1}}\Phi_N({\bf r})\,+\,\Psi_N({\bf r})\right)+\frac{\eta}{12}\left(\frac{m(\hat\phi)}{T}\right)^2\partial_i\left(\vphantom{\frac{1}{1}}\Phi_N({\bf r})\,+\,\frac{5}{3}\,\Psi_N({\bf r})\right)\\
\ 
\frac{\langle T^i_{\;j}(\eta,{\bf r})\rangle_{T}}{T^4}\,&{\simeq}&\,-\frac{2\,\pi^2\,\eta^2}{225}\partial_i\partial_j\left(\vphantom{\frac{1}{1}}\Phi_N({\bf r})\,+\,\Psi_N({\bf r})\right)\,+\,\frac{\eta^2}{36}\left(\frac{m(\hat\phi)}{T}\right)^2\partial_i\partial_j\left(\vphantom{\frac{1}{1}}\Phi_N({\bf r})\,+\,\frac{7}{5}\Psi_N({\bf r})\right)\hspace{0,5cm} i\neq j\nonumber\\ \ \\
\frac{\langle T^\mu_{\;\mu}(\eta,{\bf r})\rangle_{T}}{T^4}\,&\,{\simeq}\,&\,\frac{1}{12}\left(\frac{m(\hat\phi)}{T}\right)^2\left[1\,+\,2\,\Psi_N({\bf r})\vphantom{\frac{1}{1}}\right],
\end{eqnarray}
and on the lightcone ($r=|\eta|$) the results are
\begin{eqnarray}
\frac{\rho_T}{T^4}\,=\,\frac{\langle T^0_{\;0}(\eta,{\bf r})\rangle_{T}}{T^4}\,&\,{\simeq}\,&\,\frac{\pi^2}{30}\left(1\,+\,4\,\Psi_N({\bf r})\vphantom{\frac{1}{1}}\right)\,-\frac{1}{24}\left(\frac{m(\hat\phi)}{T}\right)^2\left[1\,-\,\Phi_N({\bf r})\,+\,5\,\Psi_N({\bf r})\vphantom{\frac{1}{1}}\right]\\
\frac{p_T}{T^4}\,=\,-\frac{\langle T^i_{\;i}(\eta,{\bf r})\rangle_{T}}{T^4}\,&\,{\simeq}\,&\,\frac{\pi^2}{90}\left( 1\,+\,4\,\Psi_N({\bf r})\vphantom{\frac{1}{1}}\right)\,-\frac{1}{24}\left(\frac{m(\hat\phi)}{T}\right)^2\left[1\,-\,\frac{1}{3}\,\Phi_N({\bf r})\,+\,3\,\Psi_N({\bf r})\vphantom{\frac{1}{1}}\right]\\
\ 
\frac{\langle T^i_{\;0}(\eta,{\bf r})\rangle_{T}}{T^4}\,&{\simeq}&\,-\frac{2\,\pi^2\,\eta}{45}\,\partial_i\left(\vphantom{\frac{1}{1}}\Phi_N({\bf r})\,+\,\Psi_N({\bf r})\right)+\frac{\eta}{12}\left(\frac{m(\hat\phi)}{T}\right)^2\partial_i\left(\vphantom{\frac{1}{1}}\Phi_N({\bf r})\,+\,\frac{5}{3}\,\Psi_N({\bf r})\right)\\
\ 
\frac{\langle T^i_{\;j}(\eta,{\bf r})\rangle_{T}}{T^4}\,&{\simeq}&\,-\frac{2\,\pi^2\,\eta^2}{225}\partial_i\partial_j\left(\vphantom{\frac{1}{1}}\Phi_N({\bf r})\,+\,\Psi_N({\bf r})\right)\,+\,\frac{\eta^2}{36}\left(\frac{m(\hat\phi)}{T}\right)^2\partial_i\partial_j\left(\vphantom{\frac{1}{1}}\Phi_N({\bf r})\,+\,\frac{7}{5}\Psi_N({\bf r})\right)\hspace{0,5cm} i\neq j\nonumber\\ \ \\
\frac{\langle T^\mu_{\;\mu}(\eta,{\bf r})\rangle_{T}}{T^4}\,&\,{\simeq}\,&\,\frac{1}{12}\left(\frac{m(\hat\phi)}{T}\right)^2\left[1\,+\,2\,\Psi_N({\bf r})\vphantom{\frac{1}{1}}\right]\,.
\end{eqnarray}
In the static limit, the energy density and pressure are
\begin{eqnarray}
\frac{\rho_T}{T^4}\,=\,\frac{\langle T^0_{\;0}(\eta,{\bf p})\rangle_{T}}{T^4}\,&\,{\simeq}\,&\,\frac{\pi^2}{30}\left(\vphantom{\frac{1}{1}}1\,-\,4\,\Phi({\bf p})\right)\,-\,\frac{1}{24}\left(\frac{m(\hat\phi)}{T}\right)^2\left(\vphantom{\frac{1}{1}}1\,-\,2\,\Phi({\bf p})\right)\\
\frac{p_T}{T^4}\,=\,-\frac{\langle T^i_{\;i}(\eta,{\bf p})\rangle_{T}}{T^4}\,&\,{\simeq}\,&\,\frac{\pi^2}{90}\left(\vphantom{\frac{1}{1}}1\,-\,4\,\Phi({\bf p})\right)\,-\,\frac{1}{24}\left(\frac{m(\hat\phi)}{T}\right)^2\left(\vphantom{\frac{1}{1}}1\,-\,2\,\Phi({\bf p})\right)\\
\frac{\langle T^\mu_{\;\mu}(\eta,{\bf p})\rangle_{T}}{T^4}\,&\,{\simeq}\,&\,\frac{1}{12}\left(\frac{m(\hat\phi)}{T}\right)^2\left(\vphantom{\frac{1}{1}}1\,-\,2\,\Phi({\bf p})\right)\,,
\end{eqnarray}
the non-diagonal terms being zero. Once again, these results can be interpreted as being the corresponding energy density and pressure for a classical gas at the local Tolman temperature \eqref{Tolman} in agreement with
\cite{Nakazawa} and \cite{Holstein}. The same expressions for the static limit apply in real space (see Appendix \ref{apppoles}).

\section{Thermal shift of the effective potential minima}
\label{sec:shift}

Once the effective potential is obtained, the value of the field for which
\begin{eqnarray}
{V_{\text{eff}}}'(\hat\phi)\,=\,0
\end{eqnarray}
determines the value attained by the classical field $\hat\phi$. The inhomogeneous contributions to the effective potential will now induce a spatial dependence
on $\hat\phi$ which can be written as
\begin{eqnarray}
\hat\phi_{}(\eta,{\bf x})\,=\,\hat\phi_0+\Delta\hat\phi(\eta,{\bf x}),
\end{eqnarray}
where $\hat\phi_0$ is the minimum of the potential in the absence of metric
perturbations, but including the one-loop corrections, i.e.
\begin{eqnarray}
{V_{\text{eff}}^{\text{h}}}'(\hat\phi_0)\,=\,V'(\hat\phi_0)\,+\,{V_{1}^{\text{h}}}'(\hat\phi_0)\,+\,{V_{T}^{\text{h}}}'(\hat\phi_0)\,=\,0 ,
\end{eqnarray}
then to first order in metric perturbations and taking into account that 
$V_1^i=0$ in dimensional regularization, we get

\begin{eqnarray}
\Delta \hat\phi\,=-\,\frac{{V_{T}^{\text{i}}}'(\hat\phi_0)}{{V_{\text{eff}}^{\text{h}}}''(\hat\phi_0)}\,=\,-\frac{1}{{V_{\text{eff}}^{\text{h}}}''(\hat\phi_0)}\,\left.\frac{\text{d}m^2}{\text{d}\hat\phi}\right|_{\hat\phi=\hat\phi_0}\,\left.\frac{\text{d}V^{\text{i}}_T}{\text{d}m^2}\,\right|_{\hat\phi=\hat\phi_0}\,.
\label{minh}
\end{eqnarray}
%

Thus, the relative classical field variation is given by the temperature correction
\begin{eqnarray}
\Delta \hat\phi
=\,-\frac{V^{'''}({\hat \phi}_0)}{{V_{\text{eff}}^{\text{h}}}''(\hat\phi_0)}
\left.\frac{\text{d}V^{\text{i}}_T}{\text{d}m^2}
\right|_{\hat\phi=\hat\phi_0}\,.
\end{eqnarray}
The perturbation is therefore proportional to the third derivative of the tree-level potential, so that variations in the field expectation value are only 
generated in theories with self-interactions.

In the non-relativistic limit and in the static limit we get in Fourier space
\begin{eqnarray}
\Delta \hat\phi_T(\eta,{\bf p})\,=\,\,\frac{e^{-m(\hat\phi)/T}}{4\sqrt{2}\,\pi^{3/2}}\,\,\left(\frac{m(\hat\phi)}{T}\right)^{3/2}
V^{'''}({\hat \phi}_0)\,\left(\frac{T^2}{{V_{\text{eff}}^{\text{h}}}''(\hat\phi_0)}\right)
\,\Phi({\bf p})\,.
\end{eqnarray}
%
%
%
In the ultra-relativistic limit, we obtain for arbitrary $\eta$
\begin{eqnarray}
\Delta \hat\phi(\eta,{\bf p})\,=\,-\,\frac{V^{'''}({\hat \phi}_0)}{12}\left(\frac{T^2}{{V_{\text{eff}}^{\text{h}}}''(\hat\phi_0)}\right)\,\,\left[\left(\frac{\sin(p\,\eta)}{p\,\eta}\,-\,1\right)\Phi({\bf p})\,+\left(\frac{\sin(p\,\eta)}{p\,\eta}\right)\Psi({\bf p})\right]\,.
\end{eqnarray}
which in the static limit reduces to
\begin{eqnarray}
\Delta \hat\phi({\bf p})\,=\,\,\frac{V^{'''}({\hat \phi}_0)}{12}\left(\frac{T^2}{{V_{\text{eff}}^{\text{h}}}''(\hat\phi_0)}\right)\Phi({\bf p}),
\label{staticshift}
\end{eqnarray}
valid also in real space replacing $\Phi({\bf p})$ by $\Phi({\bf r})$. In particular, in real space we get for Newtonian potentials inside the lightcone ($r<|\eta|$)
\begin{eqnarray}
\Delta{{\hat \phi}}(\eta,r)\,=\,-\frac{\, V^{'''}({\hat \phi}_0)}{12}\left(\frac{T^2}{{V_{\text{eff}}^{\text{h}}}''(\hat\phi_0)}\right)\,\left[\left(\frac{r}{|\eta|}-1\right)\Phi_N({\bf r})\,+\,\frac{r}{|\eta|}\Psi_N({\bf r})\right]\,\,
\end{eqnarray}
while oustide and on the lightcone ($r\geq|\eta|$)
\begin{eqnarray}
\Delta{{\hat \phi}}(\eta,r)\,=\,-\frac{\, V^{'''}({\hat \phi}_0)}{12}\left(\frac{T^2}{{V_{\text{eff}}^{\text{h}}}''(\hat\phi_0)}\right)\,\Psi_N({\bf r})\,.
\end{eqnarray}
Thus, we see that outside and on the lightcone ($r\geq|\eta|$), the result
reduces to minus the static limit result \eqref{staticshift}.
Inside the lightcone ($r<|\eta|$), the thermal shift depends on time and approaches asymptotically the static case.

From these results we see that there is a negligible shift in the classical field ${\hat \phi}$ at low temperature because of the exponential suppression, however, depending on 
the form of the tree-level potential, the shift generated 
by metric perturbations in the ultra-relativistic  limit could be relevant
in certain cases. 

Now, let us focus on the critical temperature of the phase transition $T_{\text{c}}$ defined by \cite{Mukhanov}
\begin{eqnarray}
V_{\text{eff}}(\hat\phi_0+\Delta\hat{\phi})\,=\,V_{\text{eff}}(0)
\label{tc}
\end{eqnarray}
where $V_{\text{eff}}$, $\hat{\phi}_0$ and $\Delta\hat{\phi}$ depend on the temperature $T$. Expanding equation \eqref{tc} around the critical temperature in the absence of metric perturbations $T^0_{\text{c}}$, we get for the leading order
\begin{eqnarray}
V^\text{h}_{\text{eff}}(\hat\phi_0)\,=\,V^\text{h}_{\text{eff}}(0)
\end{eqnarray}
which is the definition of $T^0_{\text{c}}$. Considering the next to leading order and solving for $\delta T_{\text{c}}= T_{\text{c}}-T^0_{\text{c}}$, we obtain the following expression for the shift in the critical temperature produced by metric perturbations\footnote{To get this expression we have redefined the effective potential by adding a function of the temperature in such a way that $V^\text{h}_{\text{eff}}(0)=0$ and $\frac{\text{d}}{\text{d}T}V^\text{h}_{\text{eff}}(0)=0$ for every $T$. This does not change the dynamics of the field since the aforementioned function of the temperature does not depend on the field $\phi$}
\begin{eqnarray}
\delta T_{\text{c}}\,=\, -\left.\frac{V^{\text{i}}_{\text{eff}}(\hat{\phi}_0)}{\frac{\text{d}}{\text{d}T}\left(V^{\text{h}}_{\text{eff}}(\hat{\phi}_0)\right)}\right|_{T=T^0_{\text{c}}}\,.
\end{eqnarray}
It can be shown (see Appendix \ref{appT}) that in the static limit 
\begin{eqnarray}
\frac{V^{\text{i}}_{\text{eff}}(\hat{\phi}_0)}{\frac{\text{d}}{\text{d}T}\left(V^{\text{h}}_{\text{eff}}(\hat{\phi}_0)\right)}\,=\,-\,T\,\Phi({\bf p})
\end{eqnarray}
therefore, in that case, the shift in the critical temperature is given by 
\begin{eqnarray}
\frac{\delta T_{\text{c}}}{T^0_{\text{c}}}\,=\,\Phi({\bf p})\,.
\end{eqnarray}
i.e. once again the curvature perturbation $\Psi$ does not contribute to the shift.
\section{Conclusions}
\label{sec:con}

Considering a scalar field  at finite temperature in an inhomogeneous static space-time, we have computed the one-loop  corrections to the effective potential and to the energy-momentum tensor induced by static scalar metric perturbations around a Minkowski 
background to first order in metric perturbations. To this aim, we have applied the formalism developed in \cite{Maroto, Higgs}. In particular we have used the explicit expressions for the perturbed field modes together with the assumptions of adiabatic evolution of the field. In order to obtain analytical expressions, the non-relativistic and ultra-relativistic limits have been considered. 

In the non-relativistic limit, we obtained the corresponding expressions in the static limit and also the limits for large-scale perturbations (small $p$) or times close to the initial time. In the ultra-relativistic limit, we obtain the complete results for arbitrary $p$ and $\eta$ up to $\Od(m/T)^5$. In the 
static limit, our results agree with those in \cite{Nakazawa} and  \cite{Holstein} which were obtained by means of the Schwinger-de Witt expansion. The energy density and pressure in the static limit are consistent with a local thermal distributions at the local Tolman temperature. Besides, our results are sensitive to the initial conditions set at the initial time for the mode solutions.

We have also discussed the space-dependent shift in the classical field induced by the 
metric perturbations. As expected, in the non-relativistic limit the shift is 
Boltzmann suppressed. However, in the ultra-relativistic case and depending on the shape
of the potential, the shift could be non-negligible.  

The results of the paper have shown that mode summation is a useful technique to obtain 
explicit expressions for one-loop quantities at zero and finite temperature. Unlike 
the more standard Schwinger-de Witt expansion, this method allows to calculate not only the 
local contributions to the effective action, but also the finite non-local ones which will
appear at second order in the perturbative expansion. Future work along this line  
will allow to explore this possibility.   

\vspace{0.2cm}
{\it Acknowledgements}. This work has been supported by the Spanish MICINNs Consolider-Ingenio 2010 Programme under grant MultiDark CSD2009-00064, by the Spanish Research Agency (Agencia Estatal de Investigaci\'on) through the grant IFT Centro de Excelencia Severo Ochoa SEV-2016-0597 and MINECO grants FIS2014-52837-P, FIS2016-78859-P(AEI/FEDER, UE),  AYA-2012-31101 and AYA2014-60641-C2-1-P. FDA acknowledges financial support from `la Caixa'-Severo Ochoa doctoral fellowship.


\appendix

\section{Perturbed mode solution}
\label{appsol}
The expression for $P_k(\eta,{\bf p})$ and $\delta\theta_k(\eta,{\bf p})$ are given by \cite{Higgs} (see also \cite{ACM1,ACM2})

\begin{eqnarray}
\label{Pk}
P_k(\eta,{\bf p})\,=\,
\int_0^\eta e^{-i\,{\bf k}\cdot{\bf p}\,\beta_k(\eta,\eta')}\frac{H_k(\eta',{\bf p})}{2\omega_k(\eta')}\,\text{d}\eta'+e^{-i\,{\bf k}\cdot{\bf p}\,\beta_k(\eta,0)} P_k(0,{\bf p})\,.\\
\delta\theta_k(\eta,{\bf p})\,=\, \int_{0}^\eta e^{-i\,{\bf k}\cdot {\bf p}\,\beta_k(\eta,\eta')} \, G_k(\eta',{\bf p})
\,\text{d}\eta'+e^{-i\,{\bf k}\cdot {\bf p}\beta_k(\eta,0)}\delta\theta_k(0,{\bf p})\,.
\label{thetak}
\end{eqnarray}
where $P_k(0,{\bf p})$, $\delta\theta_k(0,{\bf p})$ are the intial conditions, and
\begin{eqnarray}
\beta_k(\eta_f,\eta_i)&\,=\,&\int_{\eta_i}^{\eta_f} \frac{\text{d}\eta'}{\omega_k(\eta')}\\
H_k(\eta,{\bf p})&\,=\,&\omega_k\, Q'_k(\eta,{\bf p})\,+\,T_k(\eta,{\bf p})\\
Q_k(\eta,{\bf p})&\,=\,&-i\frac{{\bf k}\cdot{\bf p}}{\omega_k^2}\,\delta\theta_k(\eta,{\bf p})+ 
\left[D-\frac{k^2}{\omega_k^2}\right]\Psi(\eta,{\bf p})\\ 
T_k(\eta,{\bf p})&\,=\,&p^2\,\delta\theta_k(\eta,{\bf p})
-i\,{\bf k}\cdot {\bf p}\left[\Phi(\eta,{\bf p})-(D-2)\,\Psi(\eta,{\bf p})\right]\,\\
G_k(\eta,{\bf p})&\,=\,&-\omega_k\left[\Phi(\eta,{\bf p})\,+\,\frac{k^2}{\omega_k^2}\,\Psi(\eta,{\bf p})\right]\,.
\end{eqnarray}

$P_k(0,{\bf p})$ is fixed by the orthonormalization condition of the modes while $\delta\theta_k(0,{\bf p})$ remains arbitrary.
The arbitrariness in $\delta\theta_k(0,{\bf p})$ can also 
be absorbed in a change of the lower integration limit in \eqref{thetak}. As we will see, only setting the time origin to $\eta_0\rightarrow -\infty$, which is equivalent to taking $\eta\rightarrow \infty$, corresponds to the exact static limit.

Full details about the solutions \eqref{Pk} and \eqref{thetak}, and about the orthonormalization condition are given in  \cite{Higgs}.

\section{Expansion in $p\,\eta$ for static space-times}
\label{appexp}

The following expansions have been used for the computation of the potential and the energy-momentum tensor
\begin{eqnarray}
\label{pserie}
\hat{P}_k(\eta,{\bf p})\,&=&\,\left(3-\frac{k^2}{\omega_k^2}\right)\Psi({\bf p})\,+\,\\&&\sum_{l=1}^\infty\,\frac{(-1)^l}{(2l+1)!}(p\,\eta)^{2l} \left(\frac{k}{\omega_k}\right)^{2l}\frac{1}{k^2\,\omega_k^2}\times\nonumber\\&& \left[\vphantom{\frac{1}{1}}\left(\vphantom{\frac{1}{1}}2k^4+(3+2l)\,k^2 m^2\right)\left(\vphantom{\frac{1}{1}}\Phi({\bf p})+\Psi({\bf p})\right)+\left(1+2l\right)\,m^4\,\Phi({\bf p})\right].\nonumber\\
\int^1_{-1}\text{d}\hat{x}\,\left(\frac{i\,k\,p\,\hat{x}}{\omega_k^2}\right)P_k(\eta,{\bf p})\,&=&\,i\sum_{l=1}^\infty\,\frac{(-1)^l}{(2l-1)!}(p\,\eta)^{2l}\left(\frac{k}{\omega_k}\right)^{2l}\times\\&&\left[\frac{1}{(2l+1)k^2\,\omega_k\,\eta}\left((2l+1)m^4\Phi({\bf p})+3k^4(\Phi({\bf p})+\Psi({\bf p}))+k^2m^2((2l+3)\Psi({\bf p})+2(l+2)\Phi({\bf p}))\vphantom{\frac{1}{1}}\right)\right.\nonumber\\ &&\left.-\frac{\omega_k}{(2l-1)k^2\eta}\left(m^2\Phi({\bf p})+k^2(\Phi({\bf p})+\Psi({\bf p}))\vphantom{\frac{1}{1}}\right)\right]\nonumber\\
\delta\theta_k(\eta,{\bf p})\,&=&\,\sum_{l=0}^\infty\,\frac{(-1)^{l+1}}{(l+1)!}\eta^{l+1} \left(\frac{i\,{\bf k}\cdot{\bf p}}{\omega_k}\right)^{l}\frac{1}{\omega_k}\left[m^2\,\Phi({\bf p})+k^2\,(\Phi({\bf p})+\Psi({\bf p}))\right]\\
\delta\hat\theta_k(\eta,{\bf p})\,&=&\,-2\sum_{l=0}^\infty\,\frac{(-1)^{l}}{(2l+1)!}\,(p\eta)^{2l} \left(\frac{k}{\omega_k}\right)^{2l}\,\frac{\omega_k\,\eta}{2l+1}\left[\left(\frac{m}{\omega_k}\right)^2\,\Phi({\bf p})+\left(\frac{k}{\omega_k}\right)^2\,(\Phi({\bf p})+\Psi({\bf p}))\right]\\
\int^1_{-1}\text{d}\hat{x}\,\left(\frac{i\,k\,p\,\hat{x}}{\omega_k^2}\right)\,\delta\theta_k(\eta,{\bf p})\,&=&\,2\sum_{l=1}^\infty\,\frac{(-1)^{l}}{(2l+1)!}\left(p\eta\right)^{2l}\left(\frac{k}{\omega_k}\right)^{2l}\left[\left(\frac{m}{\omega_k}\right)^2\,\Phi({\bf p})+\left(\frac{k}{\omega_k}\right)^2\,(\Phi({\bf p})+\Psi({\bf p}))\right].\\
\int^1_{-1}\text{d}\hat{x}\,\left(\frac{i\,k\,p\,\hat{x}}{\omega_k^2}\right)^2\,\delta\theta_k(\eta,{\bf p})\,&=&\,-4\sum_{l=1}^\infty\,\frac{(-1)^{l}}{(2l+1)!}\left(p\eta\right)^{2l}\left(\frac{k}{\omega_k}\right)^{2l}\frac{l}{\omega_k\,\eta}\left[\left(\frac{m}{\omega_k}\right)^2\,\Phi({\bf p})+\left(\frac{k}{\omega_k}\right)^2\,(\Phi({\bf p})+\Psi({\bf p}))\right].
\end{eqnarray}
Both $\hat{P}_k(\eta,{\bf p})$ and $\int^1_{-1}\text{d}\hat{x}\,\left(\frac{i\,k\,p\,\hat{x}}{\omega_k^2}\right)\,\delta\theta_k(\eta,{\bf p})$ are the main terms appearing in the computation, while the remaining ones can be obtained from these expressions.

\section{Multipole expansion and Fourier transform}
\label{apppoles}

\subsection{Fourier transform in three dimensions}

In this discussion we follow \cite{Fourier}. The Fourier transform of a function $f({\bf r})$ is defined as\footnote{With the usual abuse of notation for using the same label for the function and for its Fourier transform. Note the non-unitary convention (the factor $1/(2\pi)^3$ is introduced when going from Fourier space to real space).}
\begin{eqnarray}
f({\bf p})\,=\,\int \text{d}^3{\bf r}\,f({\bf r})\,e^{-i{\bf p}\cdot {\bf r}}
\end{eqnarray}
Then, the inverse transform is given by

\begin{eqnarray}
f({\bf r})\,=\,\int\frac{\text{d}^3{\bf p}}{(2\,\pi)^3} \,f({\bf p})\,e^{i{\bf p}\cdot {\bf r}}
\end{eqnarray}
We are interested in the following integrals
\begin{eqnarray}
I_{lm}({\bf p})\,&=&\,\int \text{d}^3{\bf r} \,f(r)\,Y_{lm}({\bf \hat{r}})\,e^{-i{\bf p}\cdot {\bf r}}\,\\
I_{lm}({\bf r})\,&=&\,\int \frac{\text{d}^3{\bf p}}{(2\pi)^3} \,f(p)\,Y_{lm}({\bf \hat{p}})\,e^{i{\bf p}\cdot {\bf r}}\,,
\end{eqnarray}
where $Y_{lm}({\bf \hat{x}})$ are the usual spherical harmonics.
Using the Rayleigh expansion
\begin{eqnarray}
e^{i{\bf p}\cdot {\bf r}}\,=\,\sum_{l=0}^\infty(2l+1)\,i^l\,j_l(p\,r)\,P_l({\bf \hat{p}}\cdot {\bf \hat{r}})
\end{eqnarray}
where $j_l(x)$ are spherical Bessel functions and $P_l(x)$ are the Legendre polynomials,  the addition theorem for spherical harmonics
\begin{eqnarray}
P_l({\bf \hat{p}}\cdot {\bf \hat{r}})\,=\,\frac{4\pi}{2l+1}\sum_{m=-l}^l Y_{lm}({\bf \hat{r}})\,Y^*_{lm}({\bf \hat{p}})\,\,
\end{eqnarray}
and the orthonormalization of the spherical harmonics
\begin{eqnarray}
\int\text{d}\Omega_p\,Y^*_{lm}({\bf \hat{p}})\,Y_{l'm'}({\bf \hat{p}})\,=\,\delta_{l\,l'}\,\delta_{m\,m'}\,,
\end{eqnarray}
$I_{lm}({\bf p})$ and $I_{lm}({\bf r})$ can be written as
\begin{eqnarray}
I_{lm}({\bf p})\,&=&\,4\pi\,(-i)^l\,Y_{lm}({\bf \hat{p}})\,\int_0^\infty\text{d}r\,r^{2}\,f(r)\,j_l(p\,r)\\
I_{lm}({\bf r})\,&=&\,\frac{i^l}{2\pi^2}\,Y_{lm}({\bf \hat{r}})\,\int_0^\infty\text{d}p\,p^{2}\,f(p)\,j_l(p\,r)\,.
\end{eqnarray}
\subsection{Multipole expansion in Fourier space}

An arbitrary potential generated by a finite static matter distribution $\rho({\bf x})$ can be written as a multipole expansion in spherical coordinates in the region outside the matter distribution as
\begin{eqnarray}
\Phi({\bf r})\,=\,-\frac{1}{r}\,\sum_{l=0}^\infty\,\sum_{m=-l}^{l}\,\frac{Q_{lm}}{r^l}\sqrt{\frac{4\pi}{2l+1}}\,Y_{lm} ({\bf \hat{r}})
\end{eqnarray}
where $Q_{lm}$ are  the spherical multipole moments of the mass distribution given by
\begin{eqnarray}
Q_{lm}\,=\,\int\,\rho({\bf r'})\,r'^{\,l}\,\sqrt{\frac{4\pi}{2l+1}}\,Y^*_{lm}({\bf \hat{r'}})\,\text{d}^3\bf{r'}
\end{eqnarray}
The Fourier transform of the potential is
\begin{eqnarray}
\Phi({\bf p})\,=\,-\frac{4\pi}{p^2}\,\sum_{l=0}^\infty\,\frac{(-i)^l}{(2l-1)!!}\sum_{m=-l}^{l}\,Q_{lm}\,p^l\,\sqrt{\frac{4\pi}{2l+1}}\,Y_{lm} ({\bf \hat{p}})\,.
\label{multipoleFourier}
\end{eqnarray}
where we have used the following result
\begin{eqnarray}
\lim_{\lambda\rightarrow0^+}\,\int_0^\infty\text{d}r\,r^{2}\,\frac{e^{-\lambda\,r}}{r^{(l+1)}}\,j_l(p\,r)\,=\,\frac{p^{l-2}}{(2l-1)!!}\,,
\end{eqnarray}
where we have introduced a regularizing factor $e^{-\lambda\,r}$ [which in fact it is only necessary for $l=0$, the remaining cases being convergent].

To get the results for potential and energy-momentum tensor in real space we have to compute the following integrals
\begin{eqnarray}
\frac{1}{(2\pi)^3}\int\,\Phi({\bf p})\,\frac{\sin(p\,\eta)}{(p\,\eta)^{2k+1}}\,e^{i{\bf p}\cdot{\bf r}}\,\text{d}^3{\bf p}\,&\stackrel{p'=p\eta}{=}&\,\frac{1}{\eta^3}\frac{1}{(2\pi)^3}\int\,\Phi({\bf p}'/\eta)\,\frac{\sin(p')}{(p')^{2k+1}}\,e^{i{\bf p}'\cdot{\bf r}/\eta}\,\text{d}^3{\bf p'}\,.
\\
\end{eqnarray}

Taking into account the multipole expansion of the potential in Fourier space \eqref{multipoleFourier}, it can be shown for each multipole that the integral will be proportional to
\begin{eqnarray}
\frac{1}{\eta^{l+1}}\int_0^\infty\text{d}p'\,e^{-\lambda\,p}\,p'^{2}\,p'^{l-2}\frac{\sin(p')}{(p')^{2k+1}}\,j_l(p'\,r/\eta)\,, 
\end{eqnarray}
where we have introduced a regularizing factor $e^{-\lambda\,p}$.
Since the spherical Bessel functions of the first kind are finite, in particular at the origin, we get that in the static limit $\eta\rightarrow\infty$, the integral goes to zero. The same argument applies for the integrals involving cosine functions.

%
%

\section{Next to leading terms in the ultrarrelativistic limit}
\label{appnexto}

Next to leading order corrections can be obtained by expanding the Bose-Einstein factor and performing the integration term by term. For instance, the integrals we are interested in are of the following form
\begin{eqnarray}
X^4\,\int_1^\infty\,\frac{f\left(\tilde{u}\right)}{e^{X\tilde{u}}\,-\,1}\,\text{d}\tilde{u}
\end{eqnarray}
where $\tilde{u}\,=\,u/X$ and $X=m(\hat\phi)/T$.
Using the Taylor expansion of the Bose-Einstein factor
\begin{eqnarray}
\frac{1}{e^{X\tilde{u}}-1}\,=\,\sum_{k=0}^\infty\,\frac{B_k}{k!}\,\frac{\tilde{u}^k}{\tilde{u}}\,X^{k-1},
\end{eqnarray}
the next to leading corrections in $X$ can be obtained as far as the integrals are convergent. $B_k$ are the Bernoulli numbers. The function $f(\tilde{u})$ appearing in the calculations behaves as $\sim\frac{1}{\tilde{u}^3}$ in the limit $\tilde{u}\rightarrow\infty$, therefore the integrals can be performed up to $k=2$.
%
%

\section{Expression for $V^{\text{i}}_{\text{eff}}$ and $\frac{\text{d}}{\text{d}T}V^{\text{i}}_{\text{eff}}$ in the static limit}
\label{appT}

Let us define (following \cite{Mukhanov})
\begin{eqnarray}
J^{\nu}(x)\,&=&\,\int_x^\infty\frac{2\,\left(u^2-x^2\right)^{\nu/2}}{e^u-1}\,\text{d}u\\
F^{(\nu)}(X)\,&=&\,\int_0^X x^{2-\nu}\,J^\nu(x)\,\text{d}x\,.
\end{eqnarray}
Then, in the static limit we have
\begin{eqnarray}
V^{\text{h}}_{\text{eff}}\,=\,\frac{T^4}{4\,\pi^2}\,F^{(1)}\left(\frac{m(\hat{\phi})}{T}\right)
\end{eqnarray}
and
\begin{eqnarray}
V^{\text{i}}_{\text{eff}}\,=\,-\frac{T^4}{4\,\pi^2}\,\left[2\,F^{(1)}\left(\frac{m(\hat{\phi})}{T}\right)\,+\,F^{(-1)}\left(\frac{m(\hat{\phi})}{T}\right)\right]\,\Phi({\bf p})\,,
\end{eqnarray}
which can be read from the equation \eqref{effpotT}.
The derivative with respect to the temperature of the homogeneous effective potential is given by
\begin{eqnarray}
\frac{\text{d}}{\text{d}T}\left(V^{\text{h}}_{\text{eff}}(\hat{\phi}_0)\right)\,=\,\frac{T^3}{4\,\pi^2}\left[4\,F^{(1)}\left(\frac{m(\hat\phi)}{T}\right)\,-\,\left(\frac{m(\hat\phi)}{T}\right)^2\,J^{(1)}\left(\frac{m(\hat\phi)}{T}\right)\right]\,.
\label{app5}
\end{eqnarray}
The second term in the righ-hand side of the last equation can be written as
\begin{eqnarray}
\left(\frac{m(\hat\phi)}{T}\right)^2\,J^{(1)}\left(\frac{m(\hat\phi)}{T}\right)\,=\,\int_0^{m(\hat\phi)/T}\frac{\text{d}}{\text{d}x}\left(x^2\,J^{(1)}(x)\right)\,\text{d}x\,=\,2\,F^{(1)}\left(\frac{m(\hat\phi)}{T}\right)\,-\,F^{(-1)}\left(\frac{m(\hat\phi)}{T}\right)\,,
\label{app6}
\end{eqnarray}
where we have used the following property of $J^{(v)}(x)$
\begin{eqnarray}
\frac{\partial J^{(\nu)}(x)}{\partial x}=-\nu\,x\,J^{(\nu-2)}(x)\,.
\end{eqnarray} 
Therefore, equations \eqref{app5} and \eqref{app6} give us
\begin{eqnarray}
\frac{V^{\text{i}}_{\text{eff}}(\hat{\phi}_0)}{\frac{\text{d}}{\text{d}T}\left(V^{\text{h}}_{\text{eff}}(\hat{\phi}_0)\right)}\,=\,-\,T\,\Phi({\bf p})\,.
\end{eqnarray}
%
%

\twocolumngrid

\end{document}